\documentclass[12pt]{iopart}

\usepackage{bm}
\usepackage{amssymb}
\usepackage{graphicx}

\usepackage{bm}
 \usepackage{color}

\usepackage[linktoc=all,unicode=true,pdfusetitle,  bookmarks=true,bookmarksnumbered=false,bookmarksopen=false, breaklinks=true,pdfborder={0 0 0},backref=false,colorlinks=true]{hyperref}
\hypersetup{linkcolor=blue,citecolor=red}

\usepackage[hyphenbreaks]{breakurl}

\newcommand{\binom}[2]{{{#1}\choose{#2}}}

\newcommand{\X}{\theta}
\newcommand{\beq}{\begin{equation}}
\newcommand{\eeq}{\end{equation}}
\newcommand{\beqa}{\begin{eqnarray}}
\newcommand{\eeqa}{\end{eqnarray}}
\newcommand{\nn}{\nonumber\\}
\newcommand{\ex}{{\text{ex}}}

\newcommand{\text}[1]{\mathrm{#1}}
\newcommand{\rr}{\mathbf{r}}
\newcommand{\vs}{\varsigma}
\newcommand{\ann}{{\text{ann}}}
\newcommand{\que}{{\text{quen}}}
\def\bal#1\eal{\begin{align}#1\end{align}}

\newcommand{\ed}[1]{\end{document}}

\eqnobysec

\usepackage{epic}
\newcommand{\xx}{5}
\newcommand{\yy}{5}
\newcommand{\xmax}{30}
\newcommand{\ymax}{22}
\newcommand{\rtwoRoneA}{\begin{picture}(\xmax,\ymax)(-\xx,\yy)
\setlength{\unitlength}{.1mm}
\put(0,0){\circle{18}}
\put(60,0){\circle{18}}
\put(30,60){\circle*{18}}
\put(4,8){\line(1,2){22}}
\put(56,8){\line(-1,2){22}}
\end{picture}}
\newcommand{\rtwoRtwoAA}{\begin{picture}(\xmax,\ymax)(-\xx,\yy)
\setlength{\unitlength}{.1mm}
\put(0,60){\circle{18}}
\put(60,60){\circle{18}}
\put(0,0){\circle*{18}}
\put(60,0){\circle*{18}}
\put(9,0){\line(1,0){42}}
\put(0,9){\line(0,1){42}}
\put(60,9){\line(0,1){42}}
\end{picture}}
\newcommand{\rtwoRtwoCC}{\begin{picture}(\xmax,\ymax)(-\xx,\yy)
\setlength{\unitlength}{.1mm}
\put(0,60){\circle{18}}
\put(60,60){\circle{18}}
\put(0,0){\circle*{18}}
\put(60,0){\circle*{18}}
\put(9,0){\line(1,0){42}}
\put(0,9){\line(0,1){42}}
\put(60,9){\line(0,1){42}}
\put(7,7){\line(1,1) {46.5}}
\end{picture}}
\newcommand{\stwoStwoAB}{\begin{picture}(\xmax,\ymax)(-\xx,\yy)
\setlength{\unitlength}{.1mm}
\put(0,60){\circle*{18}}
\put(60,60){\circle{18}}
\put(0,0){\circle{18}}
\put(60,0){\circle*{18}}
\put(9,60){\line(1,0){42}}
\put(9,0){\line(1,0){42}}
\put(0,9){\line(0,1){42}}
\put(60,9){\line(0,1){42}}
\end{picture}}
\newcommand{\stwoStwoBC}{\begin{picture}(\xmax,\ymax)(-\xx,\yy)
\setlength{\unitlength}{.1mm}
\put(0,60){\circle*{18}}
\put(60,60){\circle{18}}
\put(0,0){\circle{18}}
\put(60,0){\circle*{18}}
\put(9,60){\line(1,0){42}}
\put(9,0){\line(1,0){42}}
\put(0,9){\line(0,1){42}}
\put(60,9){\line(0,1){42}}
\put(7,53){\line(1,-1) {46.5}}
\end{picture}}

\begin{document}

\title[One-dimensional Janus fluids]{One-dimensional Janus fluids. Exact solution and mapping from the quenched to the annealed system}
\author{M A G Maestre$^1$ and A Santos$^2$}
\
\address{$^1$Departamento de Ingenier\'ia Qu\'imica y Qu\'imica F\'isica, Universidad de
Extremadura, E--06006 Badajoz, Spain\\
$^2$Departamento de F\'isica and Instituto de Computaci\'on Cient\'ifica Avanzada (ICCAEx), Universidad de
Extremadura, E--06006 Badajoz, Spain}

\eads{\mailto{maestre@unex.es}, \mailto{andres@unex.es}}

\date{\today}
\begin{abstract}
{The equilibrium properties of a Janus fluid confined to a one-dimensional channel are exactly derived. The fluid is made of particles with two faces (active and passive), so that the pair interaction is that of hard spheres, except if the two active faces are in front of each other, in which case the interaction has a square-well attractive tail. Our exact solution refers to \emph{quenched} systems (i.e., each particle has a fixed face orientation), but we argue by means of statistical--mechanical tools that the results also apply  to \emph{annealed} systems (i.e., each particle can flip its orientation) in the thermodynamic limit. Comparison between  theoretical results and Monte Carlo simulations for quenched and annealed systems, respectively, shows an excellent agreement.}
\end{abstract}

 \maketitle

\section{Introduction\label{sec0}}

The study of colloidal particles has been a subject of increasing interest in the last decades, not only due to the numerous technological applications associated with these systems, but also regarding their fundamental role within liquid theory \cite{L01,LT11}.

Many of the mathematical models describing colloids were linked from their very inception to the concept of patchy particles, in which uniform spheres present a surface region (called a  \emph{patch}) with an interaction pattern different from that of the rest of the surface. Simple as it is, this approach of adding a little \emph{patchy} impurity on the homogeneous chemical makeup endows the model with a good deal of rich features providing a powerful tool  to get a better understanding on complex systems aggregates (regarding both organic and inorganic molecules), colloidal hierarchical structures (micelles, vesicles, nanocomposites, polymers, etc.), and, eventually, new materials synthesis.
As a matter of fact, the science of materials has experienced a true revolution thanks to the unprecedented development of innovative techniques related to the chemical synthesis technology. This avant-garde processes are generating new sets of colloidal particles with a wide range of size, composition, and anisotropic patch structure \cite{RML05,WLZL08,WM13}.

Janus fluids are made of colloidal-size particles whose surface is divided into two symmetric regions (patches) with different chemical composition, thus presenting  different behaviors \cite{WM13,BF01,F13}. The lack of centrosymmetry inherent to the pair potential yields a dynamic and vigorous surface activity derived from its anisotropic character, in some cases up to three times more interactional than a uniform particle in the same context \cite{SGP09}.
The remarkable precision of these methods makes it possible to obtain experimental results  involving chainlike or one-dimensional arrays of Janus particles \cite{YHHD10,OTSM15}.

The aim of this paper is to contribute to the understanding of the equilibrium properties of Janus fluids by focusing on one-dimensional structures. This allows us to obtain an exact description of the thermodynamic and spatial correlation quantities. Apart from its interest to model laboratory realizations of colloidal chains \cite{YHHD10,OTSM15}, the results derived in this work can be useful as a benchmark to test approximate theories.

The remainder of this paper is organized as follows. Section \ref{sec1} presents the exact equilibrium statistical--mechanical solution of a general $m$-component mixture with nearest-neighbor interactions where the interaction potential between two adjacent particles  $\alpha$ and  $\gamma=\alpha\pm 1$ may depend on their ordering (i.e., $\gamma=\alpha-1$ versus $\gamma=\alpha+1$). The particularization to a binary mixture (but yet with arbitrary anisotropic pair interactions) is worked out in section \ref{sec2} with expressions for thermodynamic quantities (density, Gibbs free energy, chemical potentials, and internal energy) and structural properties (pair correlation functions in Laplace space) as functions of pressure, temperature, and composition. Those expressions are made more explicit in section \ref{sec3}, where the Kern--Frenkel anisotropic interaction potential \cite{KF03} is considered, the thermodynamic and structural properties being plotted for several representative cases. Moreover, an analysis in section \ref{sec3} of the asymptotic decay of the pair correlation functions  shows  the absence of a Fisher--Widom transition line (separating a region in the density--temperature plane where the decay is oscillatory from a region where the decay is monotonic) \cite{FW69}. While the decay is always oscillatory, a structural crossover line exists between a region with a large wavelength from a region with a smaller wavelength. All those results correspond to a mixture of particles with quenched orientation, but in section \ref{sec5} we provide compelling arguments on the mapping of those results onto the case of one-component Janus fluids of particles with flipping orientation (annealed system). Such an equivalence is confirmed in section \ref{sec6} by comparison between the theoretical results for quenched systems and Monte Carlo simulations for annealed systems. The paper is closed by a summary and conclusions in section \ref{sec7}.

\section{General (quenched) mixture with anisotropic interactions}
\label{sec1}
\subsection{The system}
Let us consider an $N$-particle, $m$-component   fluid mixture with
number densities $\{\rho_i;i=1,\ldots,m\}$, so that the total number density is $\rho=\sum_{i=1}^m\rho_i$ and the mole fractions are $x_i=\rho_i/\rho$. The species $i$  any given particle $\alpha$ belongs to is \emph{fixed}, and in this sense the system is said to be \emph{quenched}.
Henceforth, we will use Latin and Greek indices for species and particles, respectively.

The potential energy function of
a particle $\alpha$  (located at $\mathbf{r}_\alpha$) of species $i$ due
to the interaction with another particle $\gamma$  (located at $\mathbf{r}_\gamma$) of species $j$ will be  denoted by $\phi_{ij}(\mathbf{r}_\gamma-\mathbf{r}_\alpha)$.
According to Newton's third law, $\phi_{ij}(\mathbf{r}_\gamma-\mathbf{r}_\alpha)=\phi_{ji}(\mathbf{r}_\alpha-\mathbf{r}_\gamma)$, i.e.,  $\phi_{ij}(\mathbf{r})=\phi_{ji}(-\mathbf{r})$ for all species pairs $i,j$. On the other hand, the
interaction potential is assumed to be \emph{anisotropic} and thus one may have, in general,
$\phi_{ij}(\mathbf{r})\neq \phi_{ji}(\mathbf{r})$ if $i\neq j$.
This anisotropic character means that, in general, the potential energy of a particle $\alpha$ due to the action of another particle $\gamma$ depends not only on the distance $|\mathbf{r}_\gamma-\mathbf{r}_\alpha|$ between both particles but also on the relative orientation of $\gamma$ with respect to $\alpha$.
We will also assume that $\lim_{\mathbf{r}\to\mathbf{0}}\phi_{ij}(\mathbf{r})=\infty$ and $\lim_{{r}\to\infty}\phi_{ij}(\mathbf{r})=0$, implying that the particles are impenetrable and the interactions have a finite range.

\begin{figure}
\begin{center}
\includegraphics[width=.8\columnwidth]{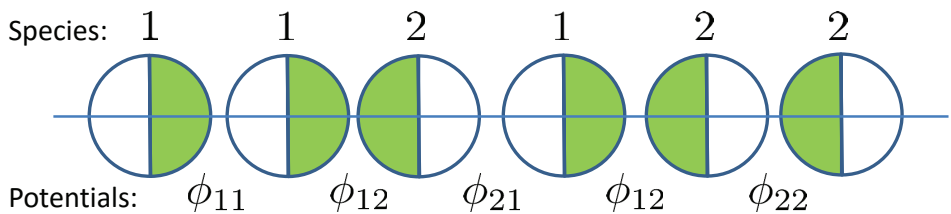}
\caption{Sketch of a binary mixture of one-dimensional Janus particles. Particles of species $1$ ($2$) have a white (green) left face and a green (white) right face. Three types of interactions are possible: green--white ($\phi_{11}$ and $\phi_{22}$), green--green ($\phi_{12}$), and white--white ($\phi_{21}$). Note that, due to invariance under reflection, one must have $\phi_{11}(r)=\phi_{22}(r)$.}
\label{sketch}
\end{center}
\end{figure}

Now we particularize to a system confined to \emph{one dimension}, so that particles are aligned along an axis of length $L$. By assuming that the interaction is restricted to nearest
neighbors, the total potential energy can be written as
\beq
\Phi_N(\mathbf{r}_1,\mathbf{r}_2,\ldots,\mathbf{r}_N)=\sum_{\alpha=1}^{N-1}\phi_{i_\alpha,i_{\alpha+1}}(\mathbf{r}_{\alpha+1}-\mathbf{r}_\alpha),
\label{N1}
\eeq
where $i_\alpha(=1,2,\ldots,m)$ denotes the species of particle $\alpha$ and, without loss of generality, we assume that particles $1,2,\ldots,N$ are ordered from left to right. Therefore, $\phi_{i_\alpha,i_{\alpha+1}}(\mathbf{r}_{\alpha+1}-\mathbf{r}_\alpha)=\phi_{i_\alpha,i_{\alpha+1}}(r_{\alpha,\alpha+1})$, where $r_{\alpha,\alpha+1}\equiv|\mathbf{r}_{\alpha+1}-\mathbf{r}_\alpha|$. The anisotropy of the interaction implies that, in general, $\phi_{ij}(r)\neq \phi_{ji}(r)$.
Figure \ref{sketch} shows a sketch of the system in the case of a binary mixture ($m=2$) of Janus particles (see section\ \ref{sec2}).
More in general, one can imagine $w$ different side faces (or `colors') and $m=w(w-1)$ species corresponding to the different ways of ordering pairs of unequal faces.
It is also possible to think of an $m$-component mixture where every particle of a given species has a patch spin vector pointing in one of $m$ possible directions; in the polydisperse limit ($m\to\infty$), the spin vector would point in any  arbitrary direction.

\subsection{Probability densities}
Let us now  use arguments similar to those conventionally used for isotropic potentials \cite{LZ71,HC04,S14,S16,FS17,MS19} to derive the structural properties of the mixture. Given a reference particle of
species  $i$, we focus on those particles to its right and denote by $p^{(\ell,+)}_{ij}(r)dr$  the (conditional) probability that its $\ell$th right neighbor belongs to species $j$ and is located at a distance between $r$ and $r+dr$. In particular, $p^{(1,+)}_{ij}(r)$ is the right
\emph{nearest-neighbor}   probability distribution function. Obviously, if $\ell=0$ one has $p^{(0,+)}_{ij}(r)=\delta_{ij}\delta(r)$.
The (conditional) \emph{total} probability density of finding a particle of
species $j$ at a distance $r$ to the right of a given particle of species $i$  is
\beq
p_{ij}^{(+)}(r)\equiv\sum_{\ell=1}^\infty p_{ij}^{(\ell,+)}(r).
\label{5.2}
\eeq
In making the upper limit of the summation equal to infinity, we are assuming the thermodynamic
limit ($L\to\infty$, $N\to\infty$, $\rho = N/L=\text{constant}$).

Let us consider now a few basic relations. First, since the $\ell$th right neighbor must be somewhere and belong to any of the species, the normalization condition reads
\beq
\sum_{j=1}^m\int_0^\infty \rmd r\, p^{(\ell,+)}_{ij}(r)=1.
\label{5}
\eeq
As before, the infinite upper limit of the integral reflects the thermodynamic limit assumption.
An obvious condition is the recurrence relation \cite{S16}
\beq
p^{(\ell,+)}_{ij}(r)=\sum_{k=1}^m\int_0^r  \rmd r' \,
p^{(\ell-1,+)}_{ik}(r')p^{(1,+)}_{kj}(r-r').
\label{7}
\eeq
Equation \eref{7} allows one to prove by induction that the normalization condition
\eref{5} is satisfied by all $p^{(\ell,+)}_{ij}(r)$, provided it is
satisfied by $p^{(1,+)}_{ij}(r)$.

Another  physical condition is \cite{LZ71,HC04,S16}
\beq
\lim_{r\to\infty}\frac{p_{ij}^{(1,+)}(r)}{p_{ik}^{(1,+)}(r)}=\mbox{independent
of $i$}.
\label{symm}
\eeq
This means that the ratio between the two probabilities that the right nearest neighbor of a given reference particle is located at a certain distance $r$ and belongs to species $j$ and $k$, respectively,  becomes asymptotically insensitive to the nature of the reference particle in the limit of large separations. This is a consequence of the fact that the reference particle and its  nearest neighbor do not interact if $r$ is beyond the range of $\phi_{ij}(r)$ or $\phi_{ik}(r)$.

In analogy with $p^{(\ell,+)}_{ij}(r)$ and $p^{(+)}_{ij}(r)$, one can introduce the distribution
$p^{(\ell,-)}_{ij}(r)$ for neighbors of species $j$ located at a distance $r$ to the left of $i$, as well as the associated total distribution
$p_{ij}^{(-)}(r)$. Obviously, the symmetry relation
\beq
\rho_ip^{(\ell,+)}_{ij}(r)=\rho_j p^{(\ell,-)}_{ji}(r)
\label{6}
\eeq
holds. Even though, in general, $ p^{(\ell,+)}_{ij}(r)\neq  p^{(\ell,-)}_{ij}(r)$ at a local level, one has
\beq
\int_0^\infty \rmd r\,p^{(\ell,+)}_{ij}(r)= \int_0^\infty \rmd r\,p^{(\ell,-)}_{ij}(r).
\label{6.1}
\eeq
This implies that, given a reference particle of species $i$, the probability that its $\ell$th neighbor (regardless of the distance) belongs to species $j$ is independent of whether the neighbor is located to the right or to the left of the reference particle.
Combination of equations \eref{6} and \eref{6.1} yields
\beq
\rho_i\int_0^\infty \rmd r\,p^{(\ell,+)}_{ij}(r)= \rho_j \int_0^\infty \rmd r\,p^{(\ell,+)}_{ji}(r).
\label{6.2}
\eeq
Because of the symmetry relation \eref{6}, henceforth we will restrict ourselves to the right probability densities $p^{(\ell,+)}_{ij}(r)$ and $p^{(+)}_{ij}(r)$.

The convolution structure of the integral in equation \eref{7} suggests the introduction of the Laplace transforms
\beq
P^{(\ell)}_{ij}(s)\equiv\int_0^\infty \rmd r\,\rme^{-sr} p^{(\ell,+)}_{ij}(r),\quad
P_{ij}(s)\equiv\int_0^\infty \rmd r\,\rme^{-sr} p^{(+)}_{ij}(r),
\label{8}
\eeq
so that equation \eref{7} becomes
\beq
P^{(\ell)}_{ij}(s)=\sum_{k=1}^m P^{(\ell-1)}_{ik}(s)P^{(1)}_{kj}(s),\quad \mathsf{P}^{(\ell)}(s)=\left[\mathsf{P}^{(1)}(s)\right]^\ell,
\label{9}
\eeq
where $\mathsf{P}^{(\ell)}(s)$ is the $m\times m$ matrix of elements
$P^{(\ell)}_{ij}(s)$. Consequently, from equation \eref{5.2} we have
\beq
\mathsf{P}(s)=\mathsf{P}^{(1)}(s)\cdot
\left[\mathsf{I}-\mathsf{P}^{(1)}(s)\right]^{-1},
\label{12}
\eeq
where $\mathsf{P}(s)$ is the matrix whose elements are $P_{ij}(s)$  and $\mathsf{I}$ is the
$m\times m$ identity matrix.

Condition \eref{5} for $\ell=1$ is
equivalent to
\beq
\sum_{j=1}^m{P}_{ij}^{(1)}(0)=1
\label{18}
\eeq
for any $i$, what implies $\det\left[\mathsf{I}-\mathsf{P}^{(1)}(0)\right]=0$.
Thus, the matrix $\mathsf{P}(s)$ is singular at $s=0$. Also, equation \eref{6.2} implies
\beq
\rho_i P_{ij}^{(\ell)}(0)=\rho_j P_{ji}^{(\ell)}(0).
\label{new6}
\eeq
As in the case of equations \eref{5}, the recursion relation \eref{9} allows one to prove by induction that equation \eref{new6} is satisfied for all $\ell$ provided it holds for $\ell=1$.

\subsection{Pair correlation function}
The probability distribution $p_{ij}^{(+)}(r)$ is related to the pair correlation
function $g_{ij}(r)$ (where a particle of species $j$ is supposed to be located at a distance $r$ to the right of a particle of species $i$) by \cite{LZ71,S16}
\beq
\rho_j g_{ij}(r)=p_{ij}^{(+)}(r).
\label{11}
\eeq
Note that, in general, $g_{ij}(r)\neq g_{ji}(r)$. One can also define an \emph{average} pair correlation function as
\beq
{g}(r)= \sum_{i,j}x_ix_j g_{ij}(r).
\label{70.2}
\eeq

In Laplace space, equation \eref{11} becomes
\beq
G_{ij}(s)=\frac{1}{\rho_j}P_{ij}(s),
\label{14}
\eeq
where $G_{ij}(s)$ is the Laplace transform of $g_{ij}(r)$.  If we
denote by $H_{ij}(s)$ the Laplace transform of the total correlation function $h_{ij}(r)\equiv
g_{ij}(r)-1$, we have
\beq
G_{ij}(s)=\frac{1}{s}+H_{ij}(s).
\label{15}
\eeq
The values $H_{ij}(0)$ are related to the isothermal compressibility (see below)
and must be finite. Therefore, the behavior of $G_{ij}(s)$ for
small $s$ is
\beq
G_{ij}(s)=\frac{1}{s}+H_{ij}(0)+\mathcal{O}(s).
\label{16}
\eeq
According to equation \eref{14}, this implies
\beq
\lim_{s\to 0}s P_{ij}(s)=\rho_j.
\label{19}
\eeq
This confirms that, as said before, the matrix $\mathsf{P}(s)$ is singular at $s=0$.
Equations \eref{symm}, \eref{18}, \eref{new6}, and \eref{19} are  basic constraints on
$P_{ij}^{(1)}(s)$ that will be used later on.

\subsection{Nearest-neighbor distribution. isothermal--isobaric ensemble}
{}From equations \eref{12} and \eref{14} we see that the knowledge of the nearest-neighbor distribution functions $\{P_{ij}^{(1)}(s)\}$ suffices to determine the pair correlation functions $\{G_{ij}(s)\}$.

In the isothermal--isobaric ensemble, the $N$-body probability distribution function in configuration
space is proportional to $\rme^{-\beta p L-\beta\Phi_N(\mathbf{r}_1,\ldots,\mathbf{r}_N)}$, where $p$ is the pressure and $\beta\equiv 1/k_{\text{B}}T$ ($k_{\text{B}}$ and $T$ being the Boltzmann constant and the absolute temperature, respectively) \cite{S16}. Therefore, the evaluation of any physical quantity
implies integrating over the system size $L$ and over the particle coordinates. Thus, in this ensemble the nearest-neighbor probability distribution function is \cite{HC04,S16}
  \beq
  p^{(1,+)}_{ij}(r)\propto \int_r^\infty \rmd L \,\rme^{-\beta p L}\int_{r_2}^L \rmd r_3 \int_{r_3}^L \rmd r_4 \cdots \int_{r_{N-1}}^L\rmd r_N \,
    \rme^{-\beta\Phi_N(\mathbf{r}_1,\ldots,\mathbf{r}_N)},
    \label{6.8}
    \eeq
where, without loss of generality, we have chosen the particles $\alpha=1$ (at $r_1=0$) and $\alpha=2$ (at $r_2=r$) as the canonical nearest-neighbor pair of species $i$ and $j$, respectively.
After taking into account equation \eref{N1}, applying periodic boundary conditions, and performing the change of variables $r_\alpha\to \hat{r}_\alpha=r_\alpha-r_{\alpha-1}$ ($\alpha=3,\ldots, N$), one gets \cite{S16}
\beq
p^{(1,+)}_{ij}(r)=x_j K_{ij}\rme^{-\beta p r-\beta\phi_{ij}(r)},
\label{32b}
\eeq
where the amplitudes $K_{ij}$ are normalization constants. These $m^2$ parameters can be determined by application of the consistency conditions \eref{symm}, \eref{18}, and \eref{new6}.
Once determined, the equation of state relating $\rho$, $p$, and $T$ is obtained by application of equation \eref{19}.

First, we note that, according to equation \eref{symm}, the ratio $K_{ij}/K_{ik}$ does not depend on the index $i$. In particular, $K_{ij}/K_{ii}=K_{jj}/K_{ji}$, i.e.,
\beq
K_{ij}K_{ji}=K_{ii}K_{jj}
\label{42}
\eeq
for all pairs $(i,j)$.
Next, Laplace transformation of equation \eref{32b} yields
\beq
P_{ij}^{(1)}(s)=x_j K_{ij}\Omega_{ij}(s+\beta p),
\label{35}
\eeq
where
\beq
\Omega_{ij}(s)\equiv\int_0^\infty \rmd r\, \rme^{-sr}\rme^{-\beta\phi_{ij}(r)}.
\label{35.1}
\eeq
Note that these functions depend parameterically  on temperature.
For small $s$,
\beq
P_{ij}^{(1)}(s)=x_j K_{ij}\left[\Omega_{ij}(\beta p)+\Omega_{ij}'(\beta p)s+\mathcal{O}(s^2)\right],
\label{35.2}
\eeq
where  the prime denotes a derivative with respect to $s$.
Therefore, the normalization condition \eref{18} implies
\beq
\sum_{j=1}^m x_j K_{ij}\Omega_{ij}(\beta p)=1.
\label{norm}
\eeq
Finally, equation \eref{new6} with $\ell=1$ gives
\beq
K_{ij}\Omega_{ij}(\beta p)=K_{ji}\Omega_{ji}(\beta p).
\label{norm2}
\eeq
Equations \eref{42}, \eref{norm}, and \eref{norm2} give $m(m-1)/2+m+m(m-1)/2=m^2$ constraints that allow one to determine the $m^2$ parameters $\{K_{ij}\}$ in terms of the set of mole fractions $\{x_i\}$, the temperature $T$, and the pressure $p$.
Then, the equation of state $\rho(T,p,\{x_i\})$ is given by equation \eref{19}. Although the matrix equation \eref{19} is in principle equivalent to $m^2$ scalar equations, it turns out that all of them collapse into a single scalar equation. This consistency test, that will be checked in section \ref{sec2} for the case $m=2$, is a direct consequence of the exact character of the results presented in this section.

\section{Binary anisotropic (quenched) mixture}
\label{sec2}

\subsection{Exact solution}
Let us now particularize the general scheme of section \ref{sec1} to the case of a binary mixture ($m=2$), although the interaction potentials $\phi_{ij}(r)$ will not be specified yet.
In that case, equations  \eref{42}, \eref{norm}, and \eref{norm2} become
\numparts
\beq
K_{11}K_{22}=K_{12}K_{21},
\label{42bin}
\eeq
\beq
\fl
x_1K_{11}\Omega_{11}(\beta p)+x_2K_{12}\Omega_{12}(\beta p)=1,\quad
x_1K_{21}\Omega_{21}(\beta p)+x_2K_{22}\Omega_{22}(\beta p)=1,
\label{40b}
\eeq
\beq
K_{12}\Omega_{12}(\beta p)=K_{21}\Omega_{21}(\beta p).
\label{39b}
\eeq
\endnumparts
From equations \eref{40b} and \eref{39b} it is possible to express $K_{11}$, $K_{22}$, and $K_{21}$ in terms of $K_{12}$:
\beq
\fl
K_{11}=\frac{1-x_2K_{12}\Omega_{12}(\beta p)}{x_1\Omega_{11}(\beta p)},\quad K_{22}=\frac{1-x_1K_{12}\Omega_{12}(\beta p)}{x_2\Omega_{22}(\beta p)},
\quad K_{21}=\frac{\Omega_{12}(\beta p)}{\Omega_{21}(\beta p)}K_{12}.
\label{40}
\eeq
Then, insertion of equations \eref{40} into equation \eref{42bin} gives a quadratic equation for $K_{12}$ whose physical solution is
 \beq
 K_{12}=\frac{2}{\Omega_{12}(\beta p)\left(1+\sqrt{1-4x_1x_2 R}\right)},\quad R\equiv 1-\frac{\Omega_{11}(\beta p)\Omega_{22}(\beta p)}{\Omega_{12}(\beta p)\Omega_{21}(\beta p)}.
 \label{6.33}
 \eeq

Once the normalization constants $\{K_{ij}\}$ are known in terms of $p$, $T$, $x_1$, and $x_2=1-x_1$, we can proceed to the determination of the equation of state. First,  equation \eref{12}  in the binary case gives
\numparts
\beq
P_{11}(s)=\frac{1-P_{22}^{(1)}(s)}{D(s)}-1,\quad P_{22}(s)=\frac{1-P_{11}^{(1)}(s)}{D(s)}-1,
\label{22}
\eeq
\beq
P_{12}(s)=\frac{P_{12}^{(1)}(s)}{D(s)},\quad P_{21}(s)=\frac{P_{21}^{(1)}(s)}{D(s)},
\label{25}
\eeq
\endnumparts
where
\beq
D(s)\equiv\left[1-P_{11}^{(1)}(s)\right]\left[1-P_{22}^{(1)}(s)\right]-P_{12}^{(1)}(s)P_{21}^{(1)}(s)
\label{26}
\eeq
is the determinant of $\mathsf{I}-\mathsf{P}^{(1)}(s)$.
Using equations \eref{35.2} and \eref{40b}, one can easily prove that $D(s)=D'(0) s+\mathcal{O}(s^2)$ with
\beqa
D'(0)&=&-x_1x_2K_{12}K_{21}\left[\Omega_{12}'(\beta p)\Omega_{21}(\beta p)+\Omega_{12}(\beta p)\Omega_{21}'(\beta p)\right]\nn
&&-x_1^2K_{11}K_{21}\Omega_{11}'(\beta p)\Omega_{21}(\beta p)-x_2^2K_{12}K_{22}\Omega_{12}(\beta p)\Omega_{22}'(\beta p).
\label{27}
\eeqa
Application of equation \eref{19} gives
\numparts
\beq
\rho x_1=\frac{1-x_2K_{22}\Omega_{22}(\beta p)}{D'(0)}=\frac{x_1K_{21}\Omega_{21}(\beta p)}{D'(0)},
\label{28}
\eeq
\beq
\rho x_2=\frac{1-x_1K_{11}\Omega_{11}(\beta p)}{D'(0)}=\frac{x_2K_{12}\Omega_{12}(\beta p)}{D'(0)}.
\label{29}
\eeq
\endnumparts
Equations \eref{40b} and \eref{39b} show that, out of the four equalities in equations \eref{28} and \eref{29}, only one is new, i.e., $\rho= K_{12}\Omega_{12}(\beta p)/D'(0)$.
Therefore, the equation of state is
  \beq
  \fl
    -\frac{1}{\rho(T,p,x_1)}=x_1^2 K_{11}\Omega_{11}'(\beta p)+x_2^2 K_{22}\Omega_{22}'(\beta p)+x_1x_2 \left[K_{12}\Omega_{12}'(\beta p)+K_{21}\Omega_{21}'(\beta p)\right].
    \label{6.34}
    \eeq

This closes the solution to the problem. Given the four interaction potentials $\{\phi_{ij}(r)\}$, the mole fraction $x_1$, the temperature $T$, and the pressure $p$, the normalization constants $\{K_{ij}\}$ are given by equations \eref{40} and \eref{6.33}, while the number density $\rho$ is given by equation \eref{6.34}. Then, equations \eref{14}, \eref{35}, and \eref{25}  provide the pair correlation functions $\{g_{ij}(r)\}$ in Laplace space.

\subsection{Thermodynamic properties}
Since the exact explicit solution of the statistical--mechanical problem relies upon the isothermal--isobaric ensemble, the key thermodynamic quantity is the Gibbs free energy $\mathcal{G}(T,p,N_1,N_2)$ \cite{S16}. Using the thermodynamic relation $N/\rho=(\partial \mathcal{G}/\partial p)_{T,N_1,N_2}$ in combination with equation \eref{6.34}, and after some algebra, it can be found that
\beqa
\frac{\mathcal{G}(T,p,N_1,N_2)}{Nk_{\text{B}}T}&=&x_1\ln\frac{x_1\Lambda_1}{\Omega_{11}}+x_2\ln\frac{x_2\Lambda_2}{\Omega_{22}}
-\ln\frac{1+\sqrt{1-4x_1x_2R}}{2\sqrt{1-R}}\nn
&&+|x_1-x_2|\ln\frac{|x_1-x_2|+\sqrt{1-4x_1x_2R}}{(|x_1-x_2|+1)\sqrt{1-R}},
\label{GN1N2}
\eeqa
where $\Lambda_i=h/\sqrt{2\pi m_ik_{\text{B}}T}$  is the thermal de Broglie's wavelength of species $i$ ($h$  and $m_i$ being the Planck constant and the mass of a particle of species $i$, respectively) and henceforth the absence of  arguments in  functions of $s$ ($\Omega_{ij}$, $\Omega_{ij}'$, \ldots) means that those functions are evaluated at $s=\beta p$.
{}From equation \eref{GN1N2} one can derive the chemical potential of species $i$ by means of the thermodynamic relation $\mu_i=(\partial \mathcal{G}/\partial N_i)_{T,p,N_{j\neq i}}$ as
\beqa
\beta\mu_i(T,p,x_1)&=&\ln\frac{x_i\Lambda_i}{\Omega_{ii}}-\ln\frac{1+\sqrt{1-4x_1x_2R}}{2\sqrt{1-R}}\nn
&&+\text{sgn}\left(2x_i-1\right)\ln\frac{|x_1-x_2|+\sqrt{1-4x_1x_2R}}{(|x_1-x_2|+1)\sqrt{1-R}},
\label{mualpha}
\eeqa
where the sign function is $\text{sgn}(x)=+1$ if $x>0$ and $-1$ otherwise. Notice that $\mathcal{G}=N_1\mu_1+N_2\mu_2$, as should be.

The internal energy $U$ obeys the thermodynamic relation $U=\mathcal{G}-T(\partial \mathcal{G}/\partial T)_{p,N_1,N_2}-p(\partial \mathcal{G}/\partial p)_{T,N_1,N_2}$. If $\mathcal{G}$ is seen as a function of $(\beta,\beta p,N_1,N_2$) rather than as a function of $(T,p,N_1,N_2$), it is easy to check that the previous relation is equivalent to $U=(\partial \beta \mathcal{G}/\partial \beta)_{\beta p,N_1,N_2}$.
Thus, equation \eref{GN1N2} gives
\beq
\fl
\frac{U(T,p,N_1,N_2)}{N}=\frac{k_{\text{B}}T}{2}+x_1\Upsilon_{11}+x_2\Upsilon_{22}
-x_1x_2K_{12}\Omega_{12}\left(\Upsilon_{11}+\Upsilon_{22}-\Upsilon_{12}
-\Upsilon_{21}\right),
\label{UN1N2}
\eeq
where
\beq
\Upsilon_{ij}(s)\equiv-\frac{\partial \ln \Omega_{ij}(s)}{\partial\beta}=\frac{1}{\Omega_{ij}(s)}\int_0^\infty \rmd r\, \rme^{-sr}\phi_{ij}(r)\rme^{-\beta\phi_{ij}(r)}.
\label{upsilon}
\eeq

As  tests on the exact character of the solution, it is proved in  \ref{appA} that the equation of state \eref{6.34} and the internal energy \eref{UN1N2}
are consistent with  standard (virial, compressibility, and energy) routes to derive the thermodynamic quantities from the pair correlation
functions \cite{S16,HM06}.

\section{Quenched Janus particles with Kern--Frenkel interaction}
\label{sec3}
The solution of the one-dimensional statistical--mechanical problem for an arbitrary mixture with anisotropic interactions has been developed in section \ref{sec1}. Next, the specialization to binary mixtures has allowed us in section \ref{sec2} to reach more explicit and detailed expressions. Now we go a step forward and particularize to a binary mixture of Janus particles (see figure \ref{sketch}). The Janus symmetry implies that a particle of species 1 is the specular reflection of a particle of species 2, so that only three interactions need to be fixed: $\phi_{11}(r)=\phi_{22}(r)$ (green--white), $\phi_{12}(r)$ (green--green), and $\phi_{21}(r)$ (white--white).
Moreover, we assume the Kern--Frenkel model \cite{KF03}, i.e., the (`passive') white face acts as a hard sphere (HS) of diameter $\sigma$ in front of any face (either white or green), while the (`active') green face acts as a square-well (SW) sphere of hard-core diameter $\sigma$, range $\lambda\sigma$, and well depth $\epsilon$ in front of another green face. Therefore, the precise model is
\beq
\fl
\phi_{11}(r)=\phi_{22}(r)=\phi_{21}(r)=
\left\{
\begin{array}{ll}
\infty,& r<\sigma,\\0,&r>\sigma,
\end{array}
\right.
\quad
\phi_{12}(r)=
\left\{
\begin{array}{ll}
 \infty,& r<\sigma,\\
-\epsilon,&\sigma<r<\lambda\sigma,\\
0,&r>\lambda\sigma,
\end{array}
\right.
\label{2}
\eeq
where $\lambda\leq 2$. Therefore,
\beq
\fl
\Omega_{11}(s)=\Omega_{22}(s)=\Omega_{21}(s)=
\Omega(s)=\frac{\rme^{-s}}{s},\quad \Omega_{12}(s)=(1+\X)\Omega(s)-\lambda \X\Omega(\lambda s),
\label{44}
\eeq
where we have taken $\sigma=1$ and the quantity $\X\equiv \rme^{\beta\epsilon}-1$ embodies all the dependence on temperature.

\subsection{Thermodynamic properties}
Despite the simplicity of equation \eref{44}, the equation of state  \eref{6.34} gives $\rho$ as an explicit function of $\beta p$ and $\beta$ but cannot be analytically inverted to express $\beta p$ as a function of $\rho$ and $\beta$. Actually, equation \eref{6.34} can be seen as a transcendental equation for $\beta p(\rho,\beta)$ that needs to be solved numerically. On the other hand, by inserting the virial expansion
\beq
\beta p(\rho,T)=\rho+B_2(T)\rho^2+B_3(T)\rho^3+B_4(T)\rho^4+\cdots
\label{virial}
\eeq
into equation \eref{6.34} and equating terms of the same order in both sides, one can easily obtain the virial coefficients $B_n(T)$ sequentially. In particular, the second, third, and fourth coefficients are
\numparts
\beq
\fl
B_2(T)=1-x_1x_2(\lambda-1)\X,\quad B_3(T)=1-x_1x_2(\lambda-1)\X\left[3-\lambda-(\lambda-1)\X\right],
\label{B2B3}
\eeq
\beqa
\fl
B_4(T)&=&1-x_1x_2\frac{\lambda-1}{2}\X\left\{13-8\lambda+\lambda^2-3(\lambda-1)\X\left[3-\lambda-x_1x_2(\lambda-1)
\right]\right.
\nn
\fl
&&\left.
+2(\lambda-1)^2\X^2\left(1+x_1x_2\right)\right\}.
\label{B4}
\eeqa
\endnumparts

The second virial coefficient is negative, implying a prevalence of the attraction  between green--green faces in the low-density regime, if the temperature is smaller than a certain Boyle temperature $T_{\text{B}}$, i.e., $T^*\equiv k_{\text{B}}T/\epsilon<T_{\text{B}}^*=1/\ln[1+1/x_1x_2(\lambda-1)]$. At exactly $T^*=T_{\text{B}}^*$, $B_2=0$ but $B_3(T_{\text{B}})=1/x_1x_2-(2-\lambda)>0$, so that $\beta p/\rho$ is an increasing function of $\rho$ at $T^*=T_{\text{B}}^*$.

It is instructive to consider the high-temperature and low-temperature limits. The results are
\numparts
\beq
\beta p(\rho,T)=\frac{\rho}{1-\rho}-\frac{\rho^2}{(1-\rho)^2}\rme^{-(\lambda-1)\rho/(1-\rho)}x_1x_2(\lambda-1)\beta\epsilon+\mathcal{O}(\beta^2),
\label{betapHT}
\eeq
\beq
\lim_{T^*\to 0}\frac{1}{\rho}=\frac{1+\beta p}{\beta p}-\min(x_1,x_2)\frac{\lambda-1}{\rme^{(\lambda-1)\beta p}-1}.
\label{betapLT}
\eeq
\endnumparts
Equation \eref{betapHT} displays the first high-temperature correction to the equation of state of the HS Tonks gas \cite{T36} due to the SW nature of the interaction $\phi_{12}(r)$. On the other hand, equation \eref{betapLT} shows that in the opposite zero-temperature limit the equation of state is far from trivial and cannot be analytically inverted to express $\beta p$ as a function of $\rho$.

As for the excess internal energy per particle $u_\ex=U/N-k_{\text{B}}T/2$, equation \eref{UN1N2} yields
\beq
\frac{u_\ex}{\epsilon}=-x_1x_2K_{12}(1+\theta)\frac{\rme^{-\beta p}-\rme^{-\lambda\beta p}}{\beta p}.
\label{uex1}
\eeq
In the limit of low densities,
\beq
\frac{u_\ex(\rho,T)}{\epsilon}=u_2(T)\rho+u_3(T)\rho^2+u_4(T)\rho^3+\cdots,
\label{uex2}
\eeq
where
\numparts
\beq
\fl
u_2(T)=-x_1x_2(\lambda-1)\rme^{\beta\epsilon},\quad u_3(T)=-x_1x_2\frac{\lambda-1}{2}\rme^{\beta\epsilon}\left[3-\lambda-2(\lambda-1)\X\right],
\eeq
\beqa
\fl
u_4(T)&=&-x_1x_2\frac{\lambda-1}{6}\rme^{\beta\epsilon}\left\{13-8\lambda+\lambda^2-6(\lambda-1)\X\left[3-\lambda-x_1x_2(\lambda-1)
\right]\right.\nn
\fl&&\left.+6(\lambda-1)^2\X^2\left(1+x_1x_2\right)\right\}.
\eeqa
\endnumparts
The curvature of $u_\ex(\rho,T)/\epsilon$ at $\rho=0$ is dictated by the sign of the coefficient $u_3(T)$. Thus, $u_\ex(\rho,T)/\epsilon$ is concave (convex) at $\rho=0$ if $T^*<T_u^*$ ($T^*>T_u^*$), where $T_u^*=1/\ln[(\lambda+1)/2(\lambda-1)]>T_{\text{B}}^*$.
As before, it is instructive to analyze the high-temperature and low-temperature limits. The results are
\numparts
\label{uexHT-uexLT}
\beq
\lim_{T^*\to \infty}\frac{u_\ex(\rho,T)}{\epsilon}=-x_1x_2\left[1-\rme^{-(\lambda-1)\rho/(1-\rho)}\right],
\label{uexHT}
\eeq
\beq
\lim_{T^*\to 0}\frac{u_\ex(\rho,T)}{\epsilon}=-\min(x_1,x_2).
\label{uexLT}
\eeq
\endnumparts
Equation \eref{uexHT} shows that, while $\lim_{T^*\to\infty}u_\ex(\rho,T)/k_{\text{B}}T=0$, $\lim_{T^*\to\infty}u_\ex(\rho,T)/\epsilon$ is a non-trivial finite limit. In what concerns equation \eref{uexLT}, it has a simple interpretation. Suppose that $N_1\leq N_2$. At zero temperature and any finite density, the free energy is minimized by minimizing the internal energy and this corresponds to configurations of $N_1$ pairs of the type $1$--$2$ plus $N_2-N_1$ particles of species $2$. The total internal energy is then $U=-N_1/\epsilon$, what implies $u_\ex/\epsilon=-x_1$.

It is interesting to remark that the limits $\rho\to 0$ and $T^*\to 0$ do not commute. While $\lim_{\rho\to 0}\beta p(\rho,T)/\rho=1$ and $\lim_{\rho\to 0}u_\ex(\rho,T)/\epsilon=0$ at any \emph{non-zero} temperature, equations \eref{betapLT} and \eref{uexLT} imply that $\lim_{\rho\to 0}\lim_{T^*\to 0}\beta p(\rho,T)/\rho=1-\min(x_1,x_2)$ and $\lim_{\rho\to 0}\lim_{T^*\to 0}u_\ex(\rho,T)/\epsilon=-\min(x_1,x_2)$.
Notice also that equations \eref{virial}--\eref{uexLT}  are consistent with the exact thermodynamic relation \cite{S16}
$\rho^2\left(\partial u_\ex/\partial\rho\right)_{\beta,x_1}=\left(\partial \beta p/\partial\beta\right)_{\rho,x_1}$.

\begin{figure}
\begin{center}
\includegraphics[width=.8\columnwidth]{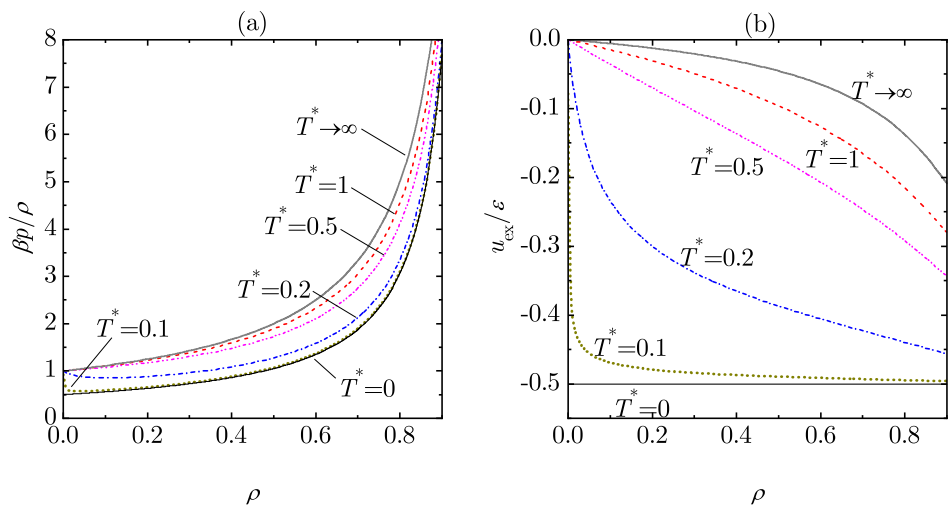}
\caption{Plot of (a) the compressibility factor $\beta p/\rho$ and (b) the excess internal energy per particle $u_\ex/\epsilon$ versus density at temperatures $T^*=0,0.1,0.2,0.5,1,\infty$ for an equimolar mixture ($x_1=x_2=\frac{1}{2}$) with $\lambda=1.2$.}
\label{fig2}
\end{center}
\end{figure}
As an illustration, figure \ref{fig2} shows $\beta p/\rho$ and $u_\ex/\epsilon$ versus $\rho$ at temperatures $T^*=0,0.1,0.2,0.5,1,\infty$ for an equimolar mixture ($x_1=x_2=\frac{1}{2}$) with $\lambda=1.2$. For such a system, $T^*_{\text{B}}=0.328$ and $T_u^*=0.587$. We observe that, as expected, the isotherms are `sandwiched' between the curves corresponding to the limits $T^*=0$ and $T^*\to\infty$. It is quite apparent that the compressibility factor $\beta p/\rho$ at $T^*=0.1$ is practically indistinguishable from the one corresponding to $T^*=0$, except for very small densities. Analogously, the isotherm $T^*=1$ in figure \ref{fig2}(a) is very close to that of infinite temperature. Thus, in contrast to what happens in the case of the conventional SW fluid \cite{S16}, the influence of temperature on the equation of state is relatively moderate. On the other hand,  temperature does play a relevant role on the excess internal energy, as figure \ref{fig2}(b) shows. A clear transition from concavity to convexity can be observed as temperature increases. In the case of $T^*=0.5<T_u^*$, although the curve is slightly concave at $\rho=0$, an inflection point is present at $\rho=0.293$, the curve becoming convex thereafter.
\subsection{Structural properties}
According to equations \eref{35} and \eref{26}, the determinant $D(s)$ becomes
\beq
D(s)=1-a\Omega(s+\beta p)-b\Omega(s+\beta p)\left[\Omega(s+\beta p)-\lambda \Omega(\lambda(s+\beta p))\right],
\label{DD}
\eeq
where we have called
\beq
a\equiv x_1K_{11}+x_2K_{22},\quad
b\equiv x_1x_2 K_{11}K_{22}\theta.
\label{ab}
\eeq
Moreover, from equations \eref{14}, \eref{22}, and \eref{25}, we find the following expressions for the pair correlation functions in Laplace space:
\numparts
\beq
G_{11}(s)=\frac{K_{11}}{\rho}\Psi^{(1,0)}(s)+\frac{x_2K_{11}K_{22}\X}{\rho}
\left[\Psi^{(2,0)}(s)-\lambda \Psi^{(1,1)}(s)\right],
\label{57}
\eeq
\beq
G_{22}(s)=\frac{K_{22}}{\rho}\Psi^{(1,0)}(s)+\frac{x_1K_{11}K_{22}\X}{\rho}
\left[\Psi^{(2,0)}(s)-\lambda \Psi^{(1,1)}(s)\right],
\label{58}
\eeq
\beq
G_{12}(s)=\frac{K_{12}(1+\X)}{\rho}\Psi^{(1,0)}(s)-\frac{K_{12}\lambda \X }{\rho}\Psi^{(0,1)}(s),
\label{59}
\eeq
\beq
G_{21}(s)=\frac{K_{21}}{\rho}\Psi^{(1,0)}(s),
\label{60}
\eeq
\endnumparts
where
\beq
\Psi^{(k_1,k_2)}(s)\equiv \frac{\left[\Omega(s+\beta p)\right]^{k_1}\left[\Omega(\lambda(s+\beta p))\right]^{k_2}}{D(s)}.
\label{Psi}
\eeq

The pair correlation functions $g_{ij}(r)$ in real space are given by expressions analogous to equations \eref{57}--\eref{60} with the replacement $\Psi^{(k_1,k_2)}(s)\to \psi^{(k_1,k_2)}(r)$, where the function $\psi^{(k_1,k_2)}(r)=\mathcal{L}^{-1}\left[\Psi^{(k_1,k_2)}(s)\right]$
is the inverse Laplace transform of $\Psi^{(k_1,k_2)}(s)$. In order to find practical representations of $g_{ij}(r)$, let us use
the mathematical identity
\beq
\left[1-ax- b x(x-y)\right]^{-1}=\sum_{n=0}^\infty\sum_{\ell=0}^{n}C_{n,\ell},
x^ny^\ell
\label{62}
\eeq
where
\beq
C_{n,\ell}\equiv \frac{a^{n-\ell} (-b)^\ell}{\ell!} \sum_{q=0}^{[(n-\ell)/2]}\frac{(n-q)!}{q!(n-\ell-2q)!}(b/a^2)^{q},
\eeq
$[(n-\ell)/2]$ denoting the integer part of $(n-\ell)/2$. Equation \eref{DD} shows that $D(s)$ has the structure $1-ax-bx(x-y)$ with $x=\Omega(s+\beta p)$ and $y=\lambda \Omega(\lambda(s+\beta p))$. Therefore,
\beqa
\Psi^{(k_1,k_2)}(s)&=&\sum_{n=0}^\infty\sum_{\ell=0}^{n}C_{n,\ell}\lambda^\ell\left[\Omega(s+\beta p)\right]^{n+k_1}
\left[\Omega(\lambda(s+\beta p))\right]^{\ell+k_2}\nn
&=&\frac{1}{\lambda^{k_2}}\sum_{n=0}^\infty\sum_{\ell=0}^{n}C_{n,\ell}
\frac{\rme^{-(n+k_1+\lambda\ell+\lambda k_2)(s+\beta p)}}{(s+\beta p)^{n+\ell+k_1+k_2}}.
\eeqa
Then, taking into account the Laplace property
\beq
\mathcal{L}^{-1}\left[\frac{\rme^{-n(s+\beta p)}}{(s+\beta p)^{\ell+1}}\right]=
\frac{(r-n)^\ell}{\ell!}\rme^{-\beta p r}\Theta(r-n),
\label{65}
\eeq
where $\Theta(x)$ is the Heaviside step function, we finally have
\beqa
\psi^{(k_1,k_2)}(r)&=&\frac{\rme^{-\beta p r}}{\lambda^{k_2}}\sum_{n=0}^\infty\sum_{\ell=0}^{n}C_{n,\ell}\frac{(r-n-k_1-\lambda\ell-\lambda k_2)^{n+\ell+k_1+k_2-1}}{(n+\ell+k_1+k_2-1)!}\nn
&&\times\Theta(r-n-k_1-\lambda\ell-\lambda k_2).
\label{psi}
\eeqa
Although in principle the summation in equation \eref{psi} extends to $n\to\infty$, truncation at $n=n_{\max}$ allows one to obtain $\psi^{(k_1,k_2)}(r)$ in the interval $1\leq r\leq n_{\max}+1+k_1+\lambda k_2$. In view of equations \eref{57}--\eref{60}, this implies that truncation at $n=n_{\max}$ guarantees the exact evaluation of $g_{ij}(r)$ up to $r=n_{\max}+2$.

\begin{figure}
\begin{center}
\includegraphics[width=.8\columnwidth]{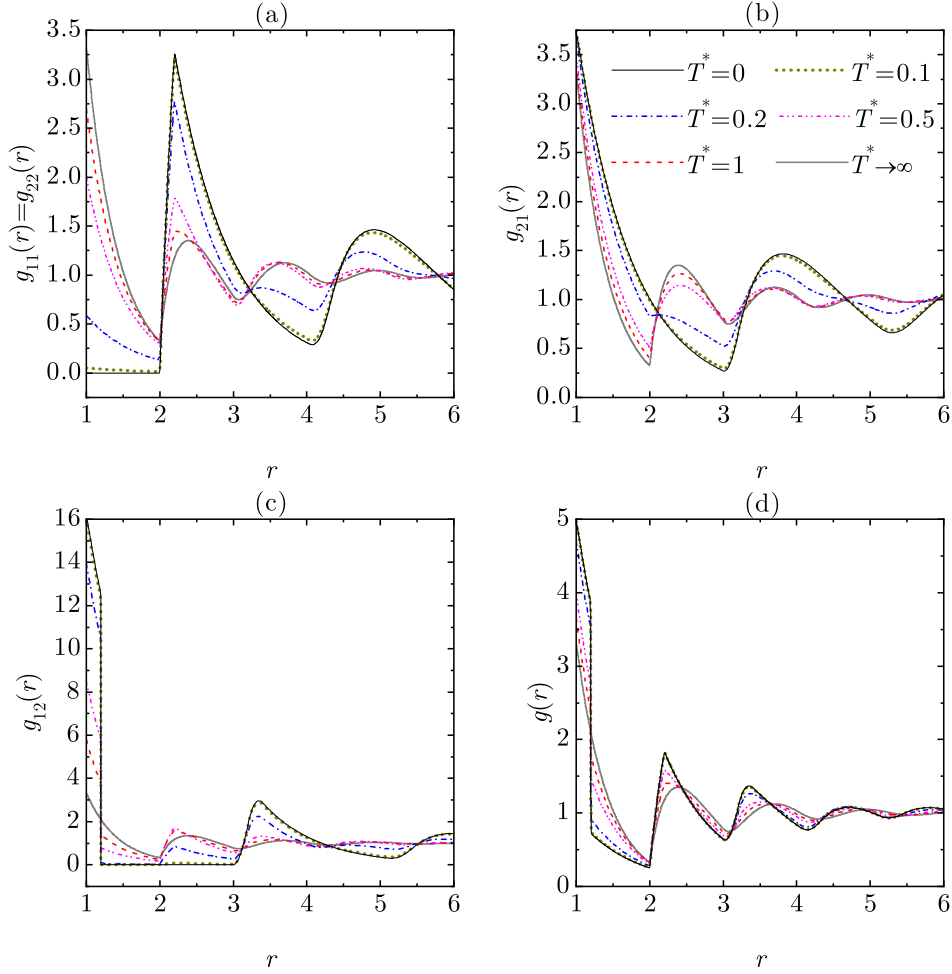}
\caption{Plot of (a) $g_{11}(r)=g_{22}(r)$, (b)  $g_{21}(r)$, (c) $g_{12}(r)$, and (d) $g(r)$  at density $\rho=0.7$ and temperatures $T^*=0,0.1,0.2,0.5,1,\infty$ for an equimolar mixture ($x_1=x_2=\frac{1}{2}$) with $\lambda=1.2$.}
\label{fig3}
\end{center}
\end{figure}

As an illustration, figure \ref{fig3} considers the same system as in figure \ref{fig2}, i.e., an equimolar mixture ($x_1=x_2=\frac{1}{2}$) with $\lambda=1.2$, and displays the pair correlation functions $g_{11}(r)=g_{22}(r)$,   $g_{21}(r)$, and $g_{12}(r)$, as well as the average function $g(r)=\frac{1}{2}g_{11}(r)+\frac{1}{4}g_{12}(r)+\frac{1}{4}g_{21}(r)$ [see equation \eref{70.2}] for the representative density $\rho=0.7$ and the same temperatures as in figure \ref{fig2}.
Clear changes in the structural properties are apparent as the temperature varies. In the limit $T^*\to\infty$, all the correlation functions are identical and coincide with that of the pure HS system. In fact, at temperature $T^*=1$ the deviations from the common HS function are rather small, except for the discontinuity of $g_{12}(r)$ and $g(r)$ at $r=\lambda$. The deviations from the HS pair correlation function become much more important as temperature decreases to $T^*=0.5$ and then to $T^*=0.2$. At $T^*=0.1$ the correlation functions are hardly distinguishable from those corresponding to $T^*=0$ and exhibit features characteristic of the expected `dimer' configurations ($1$--$2$)---($1$--$2$)---($1$--$2$)---$\cdots$, where the two particles of a dimer  ($1$--$2$) are separated a distance between $r=1$ and $r=\lambda$, while the distance between two adjacent dimers is more flexible, its typical value depending on density. Thus, we observe that, at $T^*=0$, $g_{11}(r)$ and $g_{22}(r)$ vanish for $1\leq r\leq 2$ and $g_{12}(r)$ vanish for $\lambda<r<3$.

The special sticky-hard-sphere limit \cite{S16,B68,MFGS13,FGMS13}, where $T^*\to 0$ and $\lambda\to 1$ with a constant \emph{stickiness} parameter $\tau^{-1}\equiv (\lambda-1)\theta$, is worked out in \ref{appB}.

\subsection{Asymptotic decay of correlations. Structural crossover}

The representation of $g_{ij}(r)$ in terms of the auxiliary functions $\psi^{(k_1,k_2)}(r)$ [see equations\ \eref{57}--\eref{60} and \eref{psi}] is not practical for asymptotically large values of $r$ because of the many terms involved. In that case, the asymptotic behaviors of the total correlation functions $h_{ij}(r)=g_{ij}(r)-1$ and of the average function $h(r)=g(r)-1=\sum_{i,j}x_ix_j h_{ij}(r)$ are of the form \cite{PT72}
\beq
h_{ij}(r)\sim A_{ij}\rme^{-\kappa r}\cos(\omega r+\varphi_{ij}),\quad h(r)\sim A\rme^{-\kappa r}\cos(\omega r+\varphi),
\label{h_asympt}
\eeq
where $s_\pm=-\kappa \pm \rmi \omega$ are the conjugate pair of zeroes of  $D(s)$ with a real part closest to the origin, $\rmi$ being the imaginary unit. Setting $D(s_\pm)=0$ in equation \eref{DD} yields the two coupled equations
\numparts
\beqa
\label{xi}
\rme^{-\xi}\left(\xi \cos\omega+\omega\sin\omega\right)&=&-a-\frac{b}{\xi^2+\omega^2}\left[\rme^{\lambda\xi}\left(\xi\cos\lambda\omega+\omega\sin\lambda\omega\right)\right.\nn
&&\left.-
\rme^{\xi}\left(\xi\cos\omega+\omega\sin\omega\right)\right],
\eeqa
\beqa
\label{omega}
\rme^{-\xi}\left(\xi \sin\omega-\omega\cos\omega\right)&=&\frac{b}{\xi^2+\omega^2}\left[\rme^{\lambda\xi}\left(\xi\sin\lambda\omega-\omega\cos\lambda\omega\right)\right.\nn
&&\left.-
\rme^{\xi}\left(\xi\sin\omega-\omega\cos\omega\right)\right],
\eeqa
\endnumparts
where $\xi\equiv \kappa-\beta p$.

\begin{figure}
\begin{center}
\includegraphics[width=.5\columnwidth]{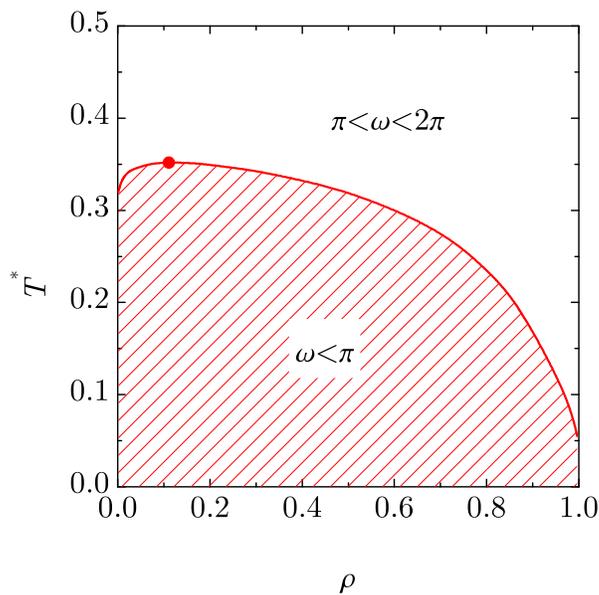}
\caption{Phase diagram for an equimolar mixture ($x_1=x_2=\frac{1}{2}$) with $\lambda=1.2$. In the states above the crossover line, the pair correlation functions present an asymptotic oscillatory behavior with a wavelength $2\pi/\omega$ comprised between the values $1$ and $2$ (i.e., a spatial frequency $\pi<\omega<2\pi$), while the wavelength $2\pi/\omega$ is larger than $2$ (i.e., $\omega<\pi$) for states below the curve. The circle represents the `critical' point $\rho_{\text{c}}=0.1105$, $T_{\text{c}}^*=0.3517$. When crossing the curve, the value of $\omega$ experiences a discontinuous change.}
\label{fig4}
\end{center}
\end{figure}

An analysis of the numerical solutions of the set of equations \eref{xi} and \eref{omega} shows that the zeroes of $D(s)$ with a real part closest to the origin are always complex numbers (i.e., $\omega\neq 0$). Therefore, no Fisher--Widom line \cite{FW69} separating the oscillatory and monotonic large-distance behaviors exists in a one-dimensional Janus fluid, in contrast to what happens in the case of one-dimensional isotropic fluids \cite{FW69,MS19,FGMS10}. Therefore, the restriction of attractive interactions to only the $1$--$2$ pair frustrates the possibility of monotonic decay of correlations, even at low temperature.

However, a structural crossover line can be identified on the plane $T^*$ vs $\rho$ separating (typically high-temperature) states where the wavelength $2\pi/\omega$ of the oscillations lies between the values $1$ and $2$  from (typically low-temperature) states with a larger wavelength. This crossover transition is reminiscent of the one observed in binary HS mixtures \cite{GDER04,SPTEP16,PBYSH20}.

The structural crossover line, and the associated phase diagram, are shown in figure \ref{fig4} for an equimolar mixture ($x_1=x_2=\frac{1}{2}$) with $\lambda=1.2$. The line has a maximum at a `critical' point $\rho_{\text{c}}=0.1105$, $T_{\text{c}}^*=0.3517$, so that if $T^*>T_{\text{c}}^*$ the asymptotic oscillatory behavior corresponds to $1<2\pi/\omega<2$. However, if $0.32< T^*<T_{\text{c}}^*$, there exists a window of densities $\rho_-(T^*)<\rho<\rho_+(T^*)$ around $\rho_{\text{c}}$ where the oscillations have a wavelength $2\pi/\omega>2$. Such a window extends to $0<\rho<\rho_+(T^*)$ if $T^*<0.32$. Upon crossing the line, the transition from shorter to longer wavelength (or vice versa) is discontinuous.

\begin{figure}
\begin{center}
\includegraphics[width=.9\columnwidth]{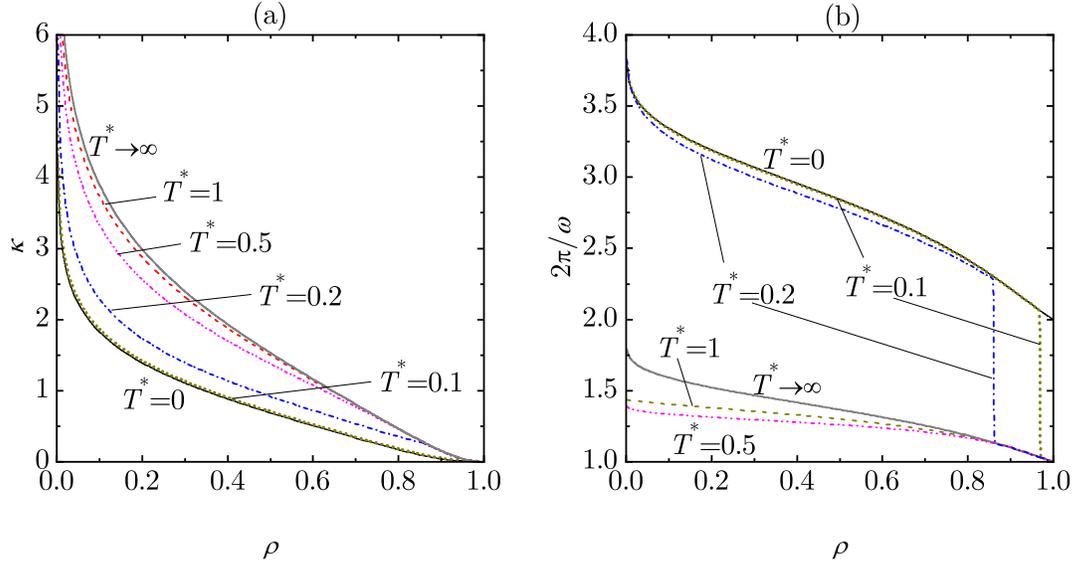}
\caption{Plot of (a) the damping coefficient $\kappa$ and (b) the wavelength  $2\pi/\omega$ versus density at temperatures $T^*=0,0.1,0.2,0.5,1,\infty$ for an equimolar mixture ($x_1=x_2=\frac{1}{2}$) with $\lambda=1.2$.}
\label{fig5}
\end{center}
\end{figure}

Figure \ref{fig5} shows $\kappa$ and $\omega$ versus $\rho$ at temperatures $T^*=0,0.1,0.2,0.5,1,\infty$ for an equimolar mixture ($x_1=x_2=\frac{1}{2}$) with $\lambda=1.2$. As density increases and/or temperature decreases, the damping coefficient $\kappa$ and  the wavelength $2\pi/\omega$ decrease. In analogy with figure \ref{fig2}(a), the curves corresponding to $T^*=0.1$ are hardly distinguishable from those corresponding to $T^*=0$, except for the discontinuous change of $\omega$ at $\rho_+(T^*=0.1)=0.969$. In the case of $T^*=0.2$ the transition takes place at $\rho_+(T^*=0.2)=0.860$.

\begin{figure}
\begin{center}
\includegraphics[width=.8\columnwidth]{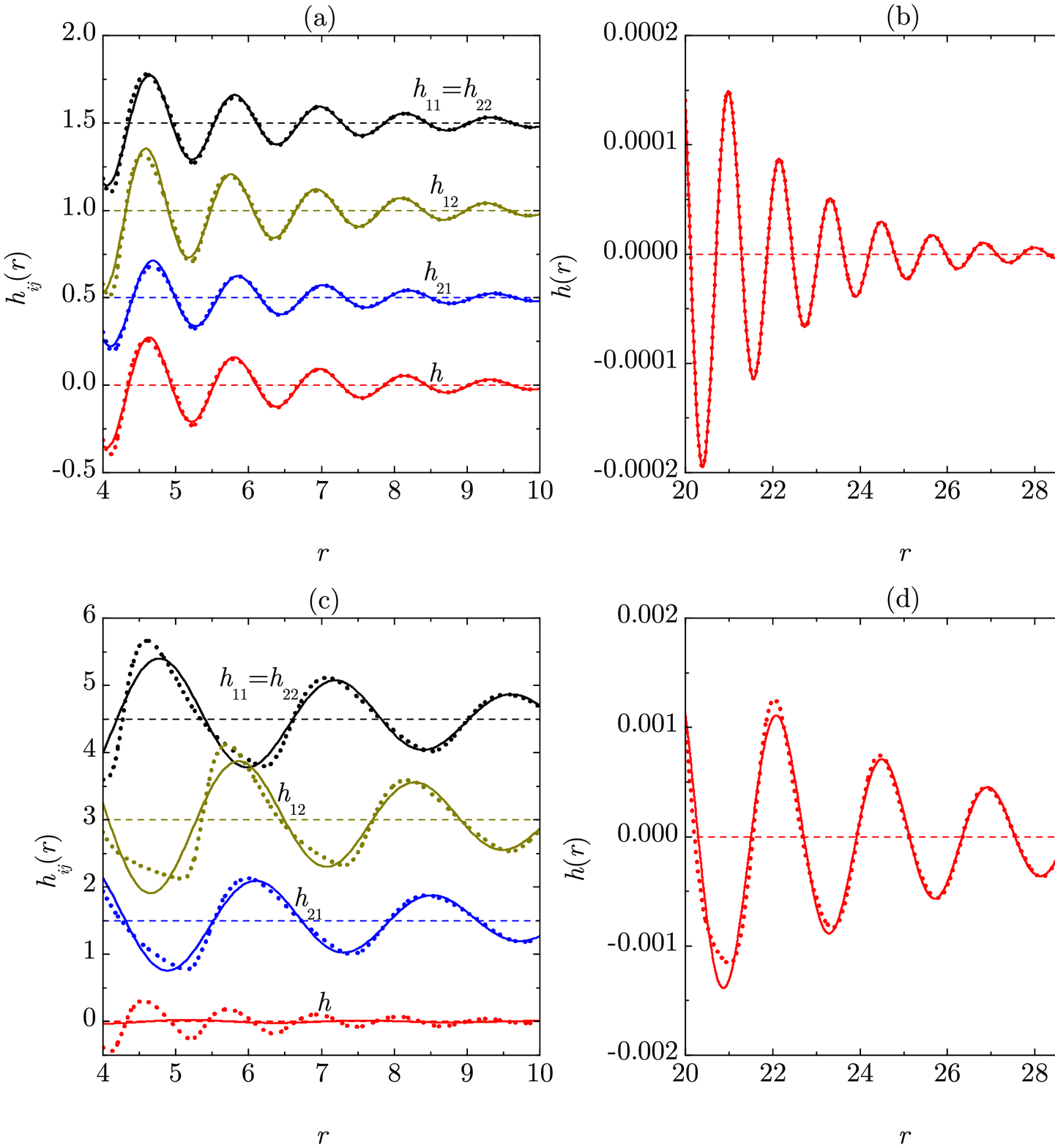}
\caption{Plot of of the total correlation functions $h_{ij}(r)$ and $h(r)$ for an equimolar mixture ($x_1=x_2=\frac{1}{2}$) with $\lambda=1.2$ at $\rho=0.8$. Panels (a) and (b) correspond to $T^*=0.5$, while panels (c) and (d) correspond to  $T^*=0.1$. The dotted curves represent the full functions, while the solid lines represent the asymptotic behavior \protect\eref{h_asympt}. Note that in panels (a) and (c) the curves representing $h_{ij}$ have been shifted vertically for better clarity.}
\label{fig6}
\end{center}
\end{figure}

The amplitudes ($A_{ij}$, $A$) and phases ($\varphi_{ij}$, $\varphi$)  in equation \eref{h_asympt} are obtained by application of the residue theorem as
\beq
A_{ij}=2|\mathcal{R}_{ij}|,\quad \varphi_{ij}=\text{arg}(\mathcal{R}_{ij}),\quad A=2|\mathcal{R}|,\quad \varphi=\text{arg}(\mathcal{R}),
\eeq
where, according to equations \eref{57}--\eref{60},
\numparts
\beq
\mathcal{R}_{11}=\frac{1-x_2 K_{22}\Omega(s_\pm+\beta p)}{\rho x_1 D'(s_\pm)},\quad
\mathcal{R}_{22}=\frac{1-x_1 K_{11}\Omega(s_\pm+\beta p)}{\rho x_2 D'(s_\pm)},
\eeq
\beq
\fl
\mathcal{R}_{12}=\frac{K_{12}\Omega_{12}(s_\pm+\beta p)}{\rho D'(s_\pm)}
,\quad \mathcal{R}_{21}=\frac{K_{21}\Omega(s_\pm+\beta p)}{\rho D'(s_\pm)}
,\quad \mathcal{R}=\sum_{i,j}x_ix_j \mathcal{R}_{ij}.
\eeq
\endnumparts

From a practical point of view, the asymptotic behavior \eref{h_asympt} is, in general, already very accurate at not too large distances.
Figure \ref{fig6} displays the partial functions $h_{ij}(r)$ and the average function $h(r)$, again for an equimolar mixture ($x_1=x_2=\frac{1}{2}$) with $\lambda=1.2$. Two representative states have been chosen, one above the transition line of figure \ref{fig4}, namely $(\rho,T^*)=(0.8,0.5)$, and the other one below the line, namely $(\rho,T^*)=(0.8,0.1)$. The associated values of the damping coefficient and the wavelength are $(\kappa,2\pi/\omega)=(0.460,1.168)$ and $(\kappa,2\pi/\omega)=(0.185,2.417)$, respectively. We observe from figure \ref{fig6}(a) that the oscillations of $h_{11}(r)=h_{22}(r)$, $h_{12}(r)$, and $h_{21}(r)$ are almost on phase, so that the average correlation function $h(r)$ is qualitatively very similar to the partial functions $h_{ij}(r)$. Figure \ref{fig6}(b) clearly shows that  $h(r)$ asymptotically oscillates with a wavelength smaller than $2$. The scenario changes at the state $(\rho,T^*)=(0.8,0.1)$. As shown by figure \ref{fig6}(c), $h_{12}(r)$ and $h_{21}(r)$ are rather on phase, but they are dephased almost half a wavelength with respect to $h_{11}(r)=h_{22}(r)$. As a consequence, a large amount of cancellation takes place when computing the average correlation function $h(r)$, its oscillations in the range $4<r<10$ being about twice as frequent as those of the partial contributions $h_{ij}(r)$. In fact, it can be seen from  figure \ref{fig6}(c) that the asymptotic function $h(r)$ corresponding to $(\kappa,2\pi/\omega)=(0.185,2.417)$ is not accurate at all in the range $4<r<10$ and one needs to move to much larger distances to match the asymptotic form, as figure \ref{fig6}(d) shows. It is interesting to note that the behavior of $h(r)$ in the range $4<r<10$ is very well captured (not shown) by the asymptotic form associated with the `competing' (subleading) root $(\kappa',2\pi/\omega')=(0.435,1.159)$.

To understand the physical origin of the crossover transition, consider the extreme cases $T^*\gg 1$ and $T^*\ll 1$. In the first case,  the asymptotic behavior of the correlation functions $h_{ij}(r)$ is qualitatively similar to that of a hard-rod one-component system, with oscillation wavelengths between the hard-core diameter and twice that value. On the other hand,  dimer-like configurations ($1$--$2$)---($1$--$2$)---($1$--$2$)---$\cdots$ prevail if $T^*\ll 1$, resulting in larger wavelengths.

\section{Orientationally constrained Janus models. Mapping of the quenched (binary-mixture) system onto the annealed (one-component) system}
\label{sec5}
In  Janus models of one-dimensional rods, only two orientations of the active face are possible, as exemplified in figure \ref{sketch}. However, in the case of particles in two and three dimensions, even if confined in a one-dimensional channel, the unit `spin' vector $\mathbf{s}_\alpha$ characterizing the orientation of the active face of particle $\alpha$ can point in any direction. In such a case, a mixture version of the model requires considering a polydisperse system with an infinite number of species, each one characterized by a different frozen spin vector.

On the other hand, one can also assume  two- or three-dimensional models where only two mutually anti-parallel (up--down) spin vectors $\mathbf{s}_+=-\mathbf{s}_-$ are allowed \cite{MFGS13,FGMS13}. In this special class of orientationally constrained Janus models, the system can be one- or two-component. In the former case (annealed system), the spin $\mathbf{s}_\alpha$ of any particle $\alpha$ is not fixed and can flip from $\mathbf{s}_\alpha=\mathbf{s}_+$ to $\mathbf{s}_\alpha=\mathbf{s}_-$, and vice versa. In contrast, in the two-component (quenched) system the spins are frozen, so that $\mathbf{s}_\alpha$ is fixed to $\mathbf{s}_\alpha=\mathbf{s}_+$ if particle $\alpha$ belongs to species $i=1$, while it is fixed to $\mathbf{s}_\alpha=\mathbf{s}_-$ if particle $\alpha$ belongs to species $i=2$. The fraction of particles having spin $\mathbf{s}_+$ \emph{fluctuates} around $\frac{1}{2}$ in the one-component (annealed) case, while it is strictly \emph{fixed} to the mole fraction $x_1=\frac{1}{2}$ in the parallel two-component (quenched) case.
More in general, in a \emph{biased} one-component system, the  fraction of particles with spin $\mathbf{s}_+$ may fluctuate around a value $x_1\neq \frac{1}{2}$. While the arguments of this section can be extended to that more general scenario, here we focus for simplicity on the unbiased case ($x_1=\frac{1}{2})$.

The total number of possible spin configurations is $2^N$ in the annealed system and $\binom{N}{N_1}\simeq \sqrt{2/\pi N}2^N$ in the quenched system with $N_1=N/2$.
The interesting question is, does there exist a relationship between the physical properties of both systems? The aim of this section is to argue that the pair correlation functions  of the annealed one-component system coincide with those of the quenched  binary mixture in the thermodynamic limit.

First, we describe the annealed and quenched versions of the system in any dimensionality.

\subsection{Annealed system}
In this case, all the particles are identical and  the specification of a microstate (in configuration space) requires, apart from the positions $\rr^N\equiv\{\rr_1,\ldots,\rr_N\}$ of the $N$ particles, the spins $\mathbf{s}^N\equiv\{\mathbf{s}_1,\ldots,\mathbf{s}_N\}$. Thus, a given microstate is specified as $\vs^N\equiv\{\vs_1,\ldots,\vs_N\}$ with the short-hand notation $\vs_\alpha\equiv \{\mathbf{s}_\alpha,\rr_\alpha\}$.
The total potential energy is
\beq
\fl
\Phi_N^\ann(\vs^N)=\sum_{\alpha=1}^{N-1}\sum_{\gamma=\alpha+1}^{N}\phi(\vs_\alpha,\vs_\gamma),
\quad
\phi(\vs_\alpha,\vs_\gamma)\equiv\phi(\mathbf{s}_\alpha,\rr_\alpha;\mathbf{s}_\gamma,\rr_\gamma) =\phi_{\mathbf{s}_\alpha,\mathbf{s}_\gamma}(\rr_{\gamma\alpha}),
\label{V.4}
\eeq
where the superscript `$\ann$' stands for `annealed.' The interaction potential $\phi(\vs_\alpha,\vs_\gamma)$ between two particles $\alpha$ and $\gamma$ depends not only on their positions $\rr_\alpha$ and $\rr_\gamma$ (actually on the relative vector $\rr_{\gamma\alpha}\equiv\rr_\gamma-\rr_\alpha$) but also on their spins $\mathbf{s}_\alpha$ and $\mathbf{s}_\gamma$

In the canonical ensemble, the probability density of the microstate $\vs^N$ is \cite{S16,HM06}
\beq
\rho_N^\ann(\vs^N)=\frac{\exp\left[-\beta \Phi_N^\ann(\vs^N)\right]}{Q_N^\ann},\quad Q_N^\ann=\int \rmd\vs^N\, \exp\left[-\beta \Phi_N^\ann(\vs^N)\right],
\label{V.3}
\eeq
where
$Q_N^\ann$ is the configuration integral and the notations $\int \rmd\vs^N\equiv \int \rmd\vs_1\cdots \int \rmd\vs_N$ and
$\int \rmd\vs_\alpha\equiv \sum_{\mathbf{s}_\alpha=\mathbf{s}_\pm}\int \rmd\rr_\alpha$
have been introduced.
The pair correlation function is defined as
\beqa
\fl
g^\ann(\vs_a,\vs_b)&=&\frac{1}{(\rho/2)^2}\left\langle \sum_{\alpha\neq \gamma}\delta(\vs_a-\vs_\alpha)\delta(\vs_b-\vs_\gamma)\right\rangle_{\rho_N^\ann}\nn
&=&\frac{4N(N-1)}{\rho^2Q_N^\ann}\int \rmd\vs^N\, \delta(\vs_a-\vs_1)\delta(\vs_b-\vs_2)\exp\left[-\beta \Phi_N^\ann(\vs^N)\right],
\label{V.10}
\eeqa
where  $\delta(\vs_a-\vs_\alpha)\equiv\delta_{\mathbf{s}_a,\mathbf{s}_\alpha}\delta(\rr_a-\rr_\alpha)$.
In the absence of interactions, $Q_N^\ann\to 2^N V^N$ and $g^\ann(\vs_a,\vs_b)\to 1$ in the thermodynamic limit.

By standard diagrammatic methods \cite{S16} one can find the virial expansion of the pair correlation function as
\beqa
\fl
g^\ann(\vs_1,\vs_2)&=&\rme^{-\beta \phi(\vs_1,\vs_2)}\left[1+\rtwoRoneA \rho+\frac{1}{2}\left(2\rtwoRtwoAA+4\rtwoRtwoCC+\stwoStwoAB+\stwoStwoBC\right)
\rho^2\right]\nn
&&+\mathcal{O}(\rho^3),
\label{V.11}
\eeqa
where, for instance,
\beqa
\fl
\rtwoRoneA =\frac{1}{2}\int \rmd\vs_3\, f(\vs_1,\vs_3)f(\vs_3,\vs_2),\quad \rtwoRtwoAA=\frac{1}{2^2}\int \rmd\vs_3\int \rmd\vs_4\, f(\vs_1,\vs_3)f(\vs_3,\vs_4)f(\vs_4,\vs_2).\nn
\label{V.12}
\eeqa
Here, $f(\vs_\alpha,\vs_\gamma)\equiv \rme^{-\beta\phi(\vs_\alpha,\vs_\gamma)}-1$ is the Mayer function.
The rest of the diagrams are defined in a similar way.

By summing over the four possible combinations $(\mathbf{s}_1,\mathbf{s}_2)$, we can finally define the radial distribution function of the annealed system as
\beq
g^\ann(r_{12})=\frac{1}{4}\sum_{\mathbf{s}_1=\mathbf{s}_\pm}\sum_{\mathbf{s}_2=\mathbf{s}_\pm}
g^\ann(\mathbf{s}_1,\rr_1;\mathbf{s}_2,\rr_2).
\eeq

\subsection{Quenched system}
Now we consider a system where $N_1$ particles  have always spin up ($\mathbf{s}_+$) and therefore belong to species $i=1$. The rest of the particles ($N_2=N-N_1$) have always spin down ($\mathbf{s}_-$) and  belong to species $i=2$. Without loss of generality we can assume that species $i=1$ is made of particles $\alpha=1,\ldots,N_1$ and species $i=2$ is made of particles $\alpha=N_1+1,\ldots,N$. The mole fractions are $x_{i}=N_{i}/N$.
In this quenched binary mixture a microstate is specified by the set of positions $\rr^N\equiv\{\rr_1,\ldots,\rr_N\}$ only, as the spins are fixed from the beginning.

In the canonical ensemble, the probability density of the microstate $\{\rr_1,\ldots,\rr_N\}$ is
\beq
\fl
\rho_{N_1,N_2}^\que(\rr^N)=\frac{\exp\left[-\beta \Phi_{N_1,N_2}^\que(\rr^N)\right]}{Q_{N_1,N_2}^\que},
\quad
Q_{N_1,N_2}^\que=\int \rmd\rr^N\, \exp\left[-\beta \Phi_{N_1,N_2}^\que(\rr^N)\right],
\label{V.13}
\eeq
where the superscript `$\que$' stands for `quenched' and
\beq
\fl
\Phi_{N_1,N_2}^\que(\rr^N)=\sum_{\alpha=1}^{N_1-1}\sum_{\gamma=\alpha+1}^{N_1}\phi_{11}(\rr_{\gamma\alpha})
+\sum_{\alpha=N_1+1}^{N-1}\sum_{\gamma=\alpha+1}^{N}\phi_{22}(\rr_{\gamma\alpha})+\sum_{\alpha=1}^{N_1}\sum_{\gamma=N_1+1}^{N}\phi_{12}(\rr_{\gamma\alpha})
\label{V.14}
\eeq
is the total potential energy. Comparison between equations \eref{V.3} and \eref{V.13} shows the relationship
\beq
Q_N^\ann=\sum_{N_1=0}^N \binom{N}{N_1}Q_{N_1,N_2}^\que.
\label{V.a}
\eeq

The three pair correlation functions of the binary mixture are defined as
\numparts
\beqa
\fl
g_{11}^\que(\rr_a,\rr_b)&=&\frac{1}{x_1^2\rho^2}\left\langle \sum_{\alpha=1}^{N_1}\sum_{\gamma=1}^{N_1}\!{'}\,\delta(\rr_a-\rr_\alpha)\delta(\rr_b-\rr_\gamma)\right\rangle_{\rho_N^\que}\nn
&=&\frac{N_1(N_1-1)}{x_1^2\rho^2Q_{N_1,N_2}^\que}\int \rmd\rr^N\, \delta(\rr_a-\rr_1)\delta(\rr_b-\rr_2)\exp\left[-\beta \Phi_{N_1,N_2}^\que(\rr^N)\right],
\label{V.16}
\eeqa
\beqa
\fl
g_{22}^\que(\rr_a,\rr_b)&=&\frac{1}{x_2^2\rho^2}\left\langle \sum_{\alpha=N_1+1}^{N}\sum_{\gamma=N_1+1}^{N}\!\!\!\!\!{'}\,\delta(\rr_a-\rr_\alpha)\delta(\rr_b-\rr_\gamma)\right\rangle_{\rho_N^\que}\nn
&=&\frac{N_2(N_2-1)}{x_2^2\rho^2Q_{N_1,N_2}^\que}\int \rmd\rr^N\, \delta(\rr_a-\rr_{N-1})\delta(\rr_b-\rr_N)\exp\left[-\beta \Phi_{N_1,N_2}^\que(\rr^N)\right],
\label{V.17}
\eeqa
\beqa
\fl
g_{12}^\que(\rr_a,\rr_b)&=&\frac{1}{x_1 x_2\rho^2}\left\langle \sum_{\alpha=1}^{N_1}\sum_{\gamma=N_1+1}^{N}\delta(\rr_a-\rr_\alpha)\delta(\rr_b-\rr_\gamma)\right\rangle\nn
&=&\frac{N_1 N_2}{x_1x_2\rho^2Q_{N_1,N_2}^\que}\int \rmd\rr^N\, \delta(\rr_a-\rr_{1})\delta(\rr_b-\rr_N)\exp\left[-\beta \Phi_{N_1,N_2}^\que(\rr^N)\right].
\label{V.18}
\eeqa
\endnumparts
In equations \eref{V.16} and \eref{V.17} the prime denotes the constraint $\alpha\neq \gamma$.
Equations \eref{V.16}--\eref{V.18} can be written in a compact way as
\beq
\fl
g_{ij}^\que(\rr_a,\rr_b)=
\frac{N_i( N_j-\delta_{ij})}{x_ix_j\rho^2Q_{N_1,N_2}^\que}\int \rmd\rr^N\, \delta(\rr_a-\rr_{I_i})\delta(\rr_b-\rr_{J_j})\exp\left[-\beta \Phi_{N_1,N_2}^\que(\rr^N)\right],
\label{V.25}
\eeq
where $I_1=1$, $I_2=N-1$, $J_1=2$, $J_2=N$.
Note that, in the absence of interactions,  $Q_{N_1,N_2}^\que=V^N$ and  $g_{ij}(\rr_a,\rr_b)\to 1$ in the thermodynamic limit.

The virial expansion of the pair correlation function $g_{ij}^\que(\rr_1,\rr_2)$ is
\beqa
\fl
g_{ij}^\que(\rr_1,\rr_2)=&\rme^{-\beta \phi_{ij}(\rr_1,\rr_2)}\left[1+\rtwoRoneA {\rho}+\frac{1}{2}\left(2\rtwoRtwoAA+4\rtwoRtwoCC+\stwoStwoAB+\stwoStwoBC\right)
\rho^2\right]\nn
&+\mathcal{O}(\rho^3),
\label{V.26}
\eeqa
where in this quenched case [compare with equation \eref{V.12} for the annealed case],
\beq
\fl
\rtwoRoneA =\int^\dagger \rmd\vs_3\, f(\vs_1,\vs_3)f(\vs_3,\vs_2),\quad \rtwoRtwoAA=\int^\dagger \rmd \vs_3\int^\dagger \rmd \vs_4\, f(\vs_1,\vs_3)f(\vs_3,\vs_4)f(\vs_4,\vs_2),
\label{V.26b}
\eeq
and so on.  Here, we have introduced the notation $\int^\dagger \rmd\vs_\alpha\equiv \sum_{i_\alpha=1,2}x_{i_\alpha}\int \rmd\rr_\alpha$,
where, in each term of the sum, particle $\alpha$ belongs to species $i_\alpha$.

In analogy with equation \eref{70.2}, the average pair correlation function is
\beq
g^\que(r_{12})=\sum_{i,j}x_ix_j g_{ij}^\que(\rr_1,\rr_2).
\eeq

\subsection{Mapping $g_{ij}^\que(\rr_1,\rr_2)\to g^\ann(\vs_1,\vs_2)$}

Comparison between equations \eref{V.10} and \eref{V.25} shows that, for a finite value of $N$ and in a strict mathematical sense,
$g^\ann(\mathbf{s}_i,\rr_1;\mathbf{s}_j,\rr_2)\neq g_{ij}^\que(\rr_1,\rr_2)$,
even if the quenched mixture is equimolar ($x_1=x_2=\frac{1}{2}$).
On the other hand, comparison between equations \eref{V.11} and \eref{V.26} shows that
\beq
g^\ann(\mathbf{s}_i,\rr_1;\mathbf{s}_j,\rr_2)= g_{ij}^\que(\rr_1,\rr_2)
\label{V.29}
\eeq
if $x_1=x_2=\frac{1}{2}$. The equality in \eref{V.29} is obvious to second order in density, as can be seen by comparison between equations \eref{V.12} and \eref{V.26b}, but extends to any order.

The solution to this paradox lies in the fact that the thermodynamic limit needs to be taken in the derivation of equations \eref{V.11} and \eref{V.26}. Thus, the equivalence between the annealed and quenched systems holds in that limit, similarly to the equivalence between different statistical ensembles. In other words, if $N\to\infty$, the huge majority of the relevant microstates in the quenched system correspond to a number of up spins practically equal to the number of down spins. As a consequence, we can expect  that [see equation \eref{V.a}]
\beq
\ln Q_N^\ann\approx \ln \left[\binom{N}{N/2}Q^\que_{N/2,N/2}\right]
\label{V.30}
\eeq
and equation \eref{V.29} holds true in the thermodynamic limit. As a plausibility argument in favor of equation \eref{V.30}, note that
\beq
\ln \sum_{N_1=0}^N\binom{N}{N_1}=N\ln 2\approx \ln \binom{N}{N/2}.
\label{V.31}
\eeq

As said before, the content of this section can easily be extended to the case where in the annealed system a certain bias makes the  fraction of particles with spin $\mathbf{s}_+$ fluctuate around a value $x_1\neq \frac{1}{2}$, in which case the equivalent quenched system has a mole fraction  $x_1\neq \frac{1}{2}$.

\section{Validation of the quenched--annealed mapping by Monte Carlo simulations}
\label{sec6}
Strictly speaking, the exact statistical--mechanical solution to the one-dimensional Janus fluid worked out in section \ref{sec3} applies to the quenched system with arbitrary composition, but not, in principle, to the annealed system. On the other hand, according to the arguments presented in section \ref{sec5}, the annealed system is expected to be described in the thermodynamic limit by the solution to the  quenched system. In order to validate and confirm this expectation, we have carried out $NVT$ Monte Carlo (MC)  simulations \cite{FS02} on the one-component, annealed system.

\begin{figure}
\begin{center}
\includegraphics[width=.8\columnwidth]{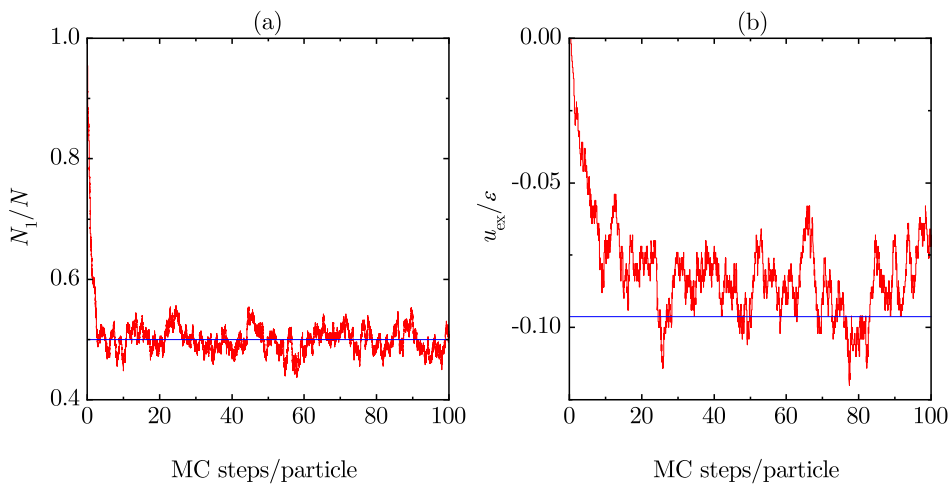}
\caption{Evolution of (a) $N_1/N$ and (b) $u_\ex/\epsilon$ versus the number of MC steps per particle for an annealed system (unbiased one-component fluid) with $\lambda=1.2$ at $T^*=1$ and $\rho=0.5$. The horizontal lines represent the equilibrium values (a) $N_1/N=\frac{1}{2}$ and (b) $u_\ex/\epsilon= -0.0962(4)$.}
\label{fig7}
\end{center}
\end{figure}

\begin{figure}
\begin{center}
\includegraphics[width=.8\columnwidth]{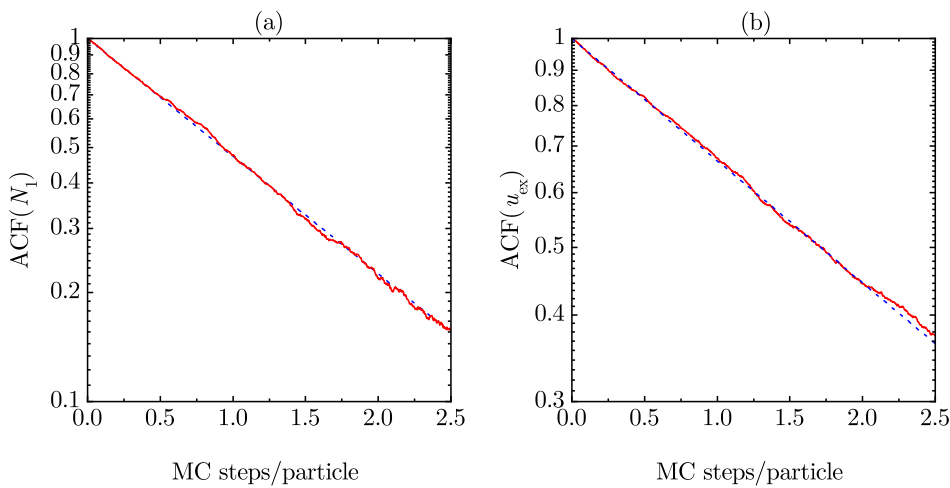}
\caption{Logarithmic plot of the ACF of (a) $N_1$ and (b) $u_\ex$ versus the number of MC steps per particle for an annealed system (unbiased one-component fluid) with $\lambda=1.2$ at $T^*=1$ and $\rho=0.5$. The dashed lines represent $\rme^{-t/t_{\text{corr}}}$, where $t$ is the number of MC steps per particle and the autocorrelation time is (a) $t_{\text{corr}}=1.34$ and (b) $t_{\text{corr}}=2.47$.}
\label{fig7b}
\end{center}
\end{figure}

In the MC simulations, a system of $N=500$ Janus particles are distributed over a ring of length $L=N/\rho$ (with periodic boundary conditions). In each  computational step, the microscopic  configuration of the system is fully determined by the position and orientation of every particle.
In order to thermalize the system and measure its equilibrium properties,  a random walk over the configuration (position plus orientation) space is performed. In each MC step, a particle is selected at random and provisionally displaced a random distance. If an overlap occurs, the displacement is rejected and a new MC step is initiated. In the absence of any overlap,  the active face of the chosen particle is provisionally assigned to its right-hand side or to its left-hand side with probabilities $q_1$ or $q_2=1-q_1$, respectively.   The attempt (displacement plus active face assignment) is accepted according to the  Metropolis criterion \cite{FS02}. The size of the position displacement is adjusted so that the acceptance ratio is approximately $50$\%. We have typically used $10^5$ MC steps per particle for equilibration plus  an additional set of $5\times 10^5$ MC steps per particle for the computation of the equilibrium quantities as averages. Except at the end of this section, we have restricted ourselves to \emph{unbiased} annealed Janus fluids ($q_1=q_2=\frac{1}{2}$).

\begin{figure}
\begin{center}
\includegraphics[width=.8\columnwidth]{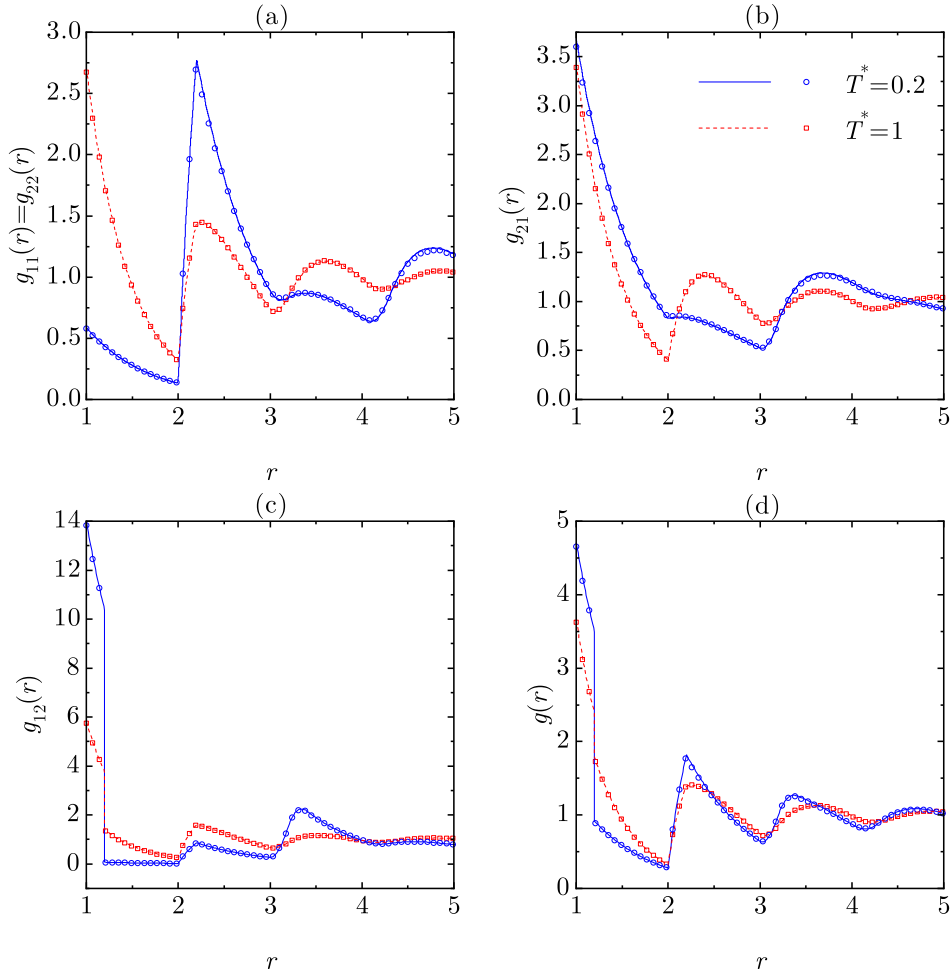}
\caption{Plot of (a) $g_{11}(r)=g_{22}(r)$, (b)  $g_{21}(r)$, (c) $g_{12}(r)$, and (d) $g(r)$  at density $\rho=0.7$ and temperatures $T^*=0.2$ and $1$ for a Janus fluid with $\lambda=1.2$. Lines represent the exact solution for the  quenched system (equimolar binary mixture), while symbols are MC results for the annealed system (unbiased one-component fluid).}
\label{fig8}
\end{center}
\end{figure}

We have simulated annealed systems with $\lambda=1.2$ at a low temperature ($T^*=0.2$) and at an intermediate temperature ($T^*=1$), in each case with densities $\rho=0.1,0.2,\ldots,0.8$.
As an illustration of the evolution of the main quantities, figure \ref{fig7} displays the MC evolution of the ratio $N_1/N$ (where $N_1$ denotes the fluctuating number of particles with a right active face) and the (reduced) excess energy per particle  $u_\ex/\epsilon$ at $T^*=1$ and $\rho=0.5$. In the initial configuration, particles are equispaced (so that $u_\ex/\epsilon=0$) with their active faces oriented to the right (so that $N_1=N$). As we can observe in figure \ref{fig7}(a), after about only $10$ MC steps per particle the fraction of particles with a given orientation fluctuates about the value $\frac{1}{2}$. On the other hand, the equilibration of the thermodynamic and structural properties is much slower. In particular, figure \ref{fig7}(b) shows that the energy has not relaxed yet to its equilibrium value after about $100$ MC steps per particle.

Figure \ref{fig7b} shows the autocorrelation function (ACF) \cite{K11,F17c} of $N_1$ and $u_\ex$ for the same case as that of figure \ref{fig7}. The ACF has been obtained in the equilibrium stage by averaging over $4000$ blocks, each one made of $2.5$ MC steps per particle.
It can be observed that in both cases the ACF decays exponentially with a characteristic autocorrelation time, $t_{\text{corr}}$, of about $1.34$ and $2.47$ MC steps per particle for $N_1$ and $u_\ex$, respectively. Therefore, the number of MC steps needed to perform sampling between statistically uncorrelated configurations in the case of energy is almost twice that in the case of the number of particles with a given orientation.

In all the simulated states, we have found an excellent agreement between the theoretical and the MC functions $g_{ij}(r)$ for the quenched and annealed systems, respectively. As an example, figure \ref{fig8} shows  $g_{11}(r)=g_{22}(r)$,   $g_{12}(r)$,  $g_{21}(r)$, and $g(r)$ at $\rho=0.7$ and $T^*=0.2$ and $1$.

\begin{figure}
\begin{center}
\includegraphics[width=.8\columnwidth]{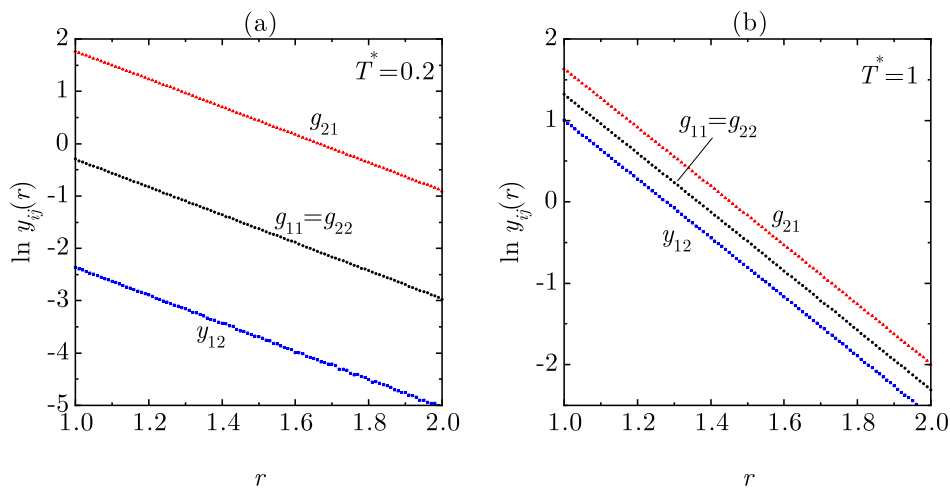}
\caption{Plot of  $\ln g_{11}(r)=\ln g_{22}(r)$, $\ln g_{21}(r)$, and $\ln y_{12}(r)$ in the first coordination shell ($1<r<2$), as obtained from MC simulations for an annealed system (unbiased one-component fluid) with $\lambda=1.2$ at $\rho=0.8$ and (a) $T^*=0.2$ and  (b)  $T^*=1$. The average slopes give (a) $\beta p=2.68$ and (b) $\beta p=3.63$, respectively.}
\label{fig9}
\end{center}
\end{figure}

Apart from the pair correlation functions, the excess internal energy per particle ($u_\ex/\epsilon$) and the pressure ($\beta p$) have been computed in the MC simulations.
While the excess internal energy can be evaluated directly, the pressure requires an alternative method. According to equations \eref{5.2}, \eref{11}, and \eref{32b}, the cavity functions
$y_{ij}(r)\equiv g_{ij}(r)\rme^{\beta\phi_{ij}(r)}$ are proportional to $\rme^{-\beta p r}$ within the first coordination shell ($1<r<2$). In the case of the Janus fluid, this means that a
logarithmic plot of $y_{11}(r)=y_{22}(r)=g_{11}(r)=g_{22}(r)$, $y_{21}(r)=g_{21}(r)$, and $y_{12}(r)=g_{12}(r)\left[1-(1-\rme^{-1/T^*})\Theta(\lambda-r)\right]$ in the region $1<r<2$ should give straight lines with a \emph{common} slope equal to $-\beta p$. As an illustration, figure \ref{fig9} shows a plot of $\ln y_{ij}(r)$ in the cases $T^*=0.2$ and $T^*=1$, both with $\rho=0.8$. It must be noted that this method to obtain the pressure cannot be applied to the average pair correlation function $g(r)$ but requires to disentangle the partial contributions  $g_{ij}(r)$.

\begin{figure}
\begin{center}
\includegraphics[width=.8\columnwidth]{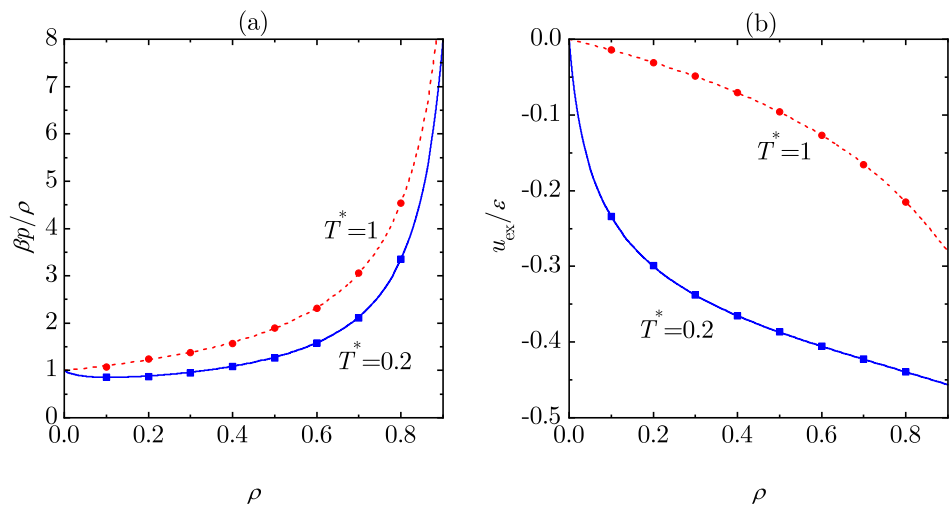}
\caption{Plot of (a) the compressibility factor $\beta p/\rho$ and (b) the excess internal energy per particle $u_\ex/\epsilon$ versus density at temperatures $T^*=0.2$ and $1$ for a Janus fluid with $\lambda=1.2$. Lines represent the exact solution for the  quenched system (equimolar binary mixture), while symbols are MC results for the annealed system (unbiased one-component fluid).}
\label{fig10}
\end{center}
\end{figure}

The thermodynamic quantities $\beta p/\rho$ and $u_\ex/\epsilon$ as functions of $\rho$ are compared with the theoretical curves for the quenched systems in figure \ref{fig10}, again with a virtually perfect agreement. We have estimated the errors in the simulation values by dividing the $5\times 10^5$ MC steps per particle into $25$ blocks, each one made of $2\times 10^4$ MC steps per particle, and checked that the error bars are smaller than the size of symbols in figure \ref{fig10}. For instance, in the state $\rho=0.5$ and $T^*=1$ we have obtained $u_\ex/\epsilon=-0.0962(4)$ and $\beta p=1.890(2)$, where the numbers enclosed by parentheses represent standard deviations. In the case of pressure, the error estimate takes into account that the linear fit of $\ln y_{ij}(r)$ is made over $100$ values equispaced between $r=1$, and $r=2$, each one with an error of about $0.014$.





\begin{figure}
\begin{center}
\includegraphics[width=.8\columnwidth]{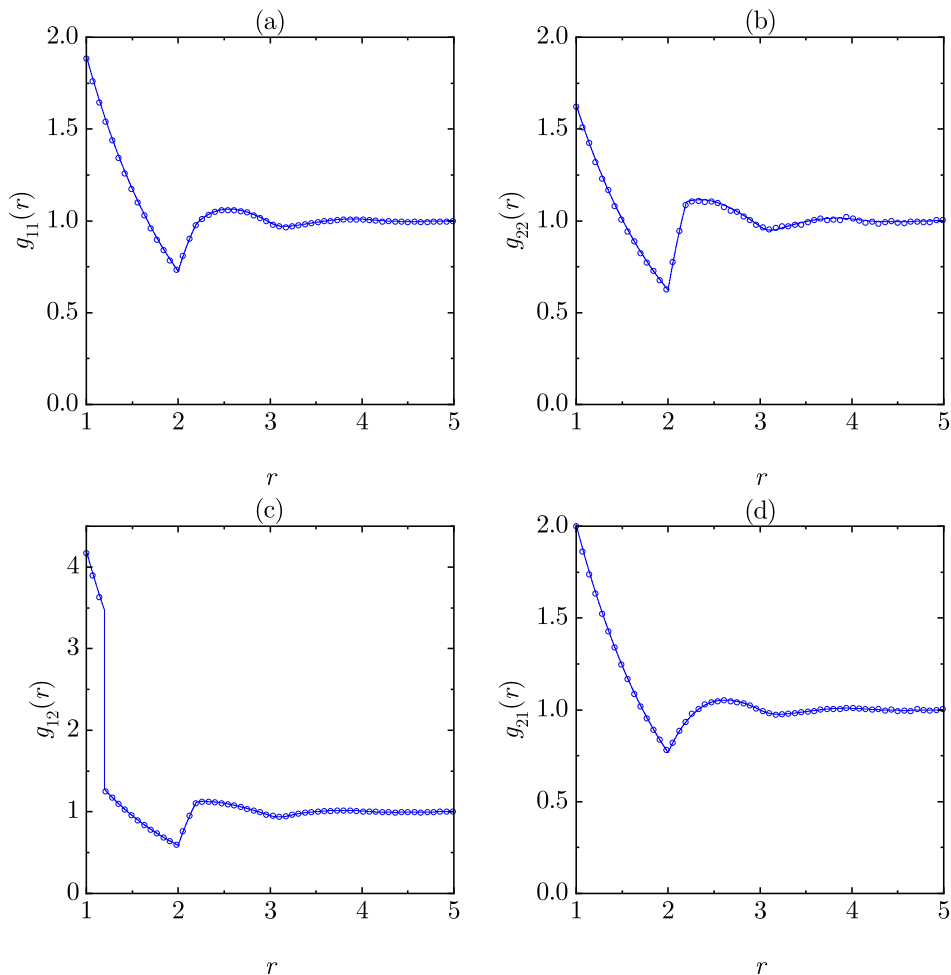}
\caption{Plot of (a) $g_{11}(r)$, (b)  $g_{22}(r)$, (c) $g_{12}(r)$, and (d) $g_{21}(r)$  at density $\rho=0.5$ and temperature $T^*=1$ for a Janus fluid with $\lambda=1.2$ and a mole fraction $x_1=0.7748$. Lines represent the exact solution for the  quenched system (nonequimolar binary mixture), while symbols are MC results for the annealed system (biased one-component fluid).}
\label{fig11}
\end{center}
\end{figure}

So far, we have paid special attention to the mapping between \emph{unbiased} annealed systems and \emph{equimolar} quenched systems. On the other hand, given that the arguments in section \ref{sec5} can be extended to the biased/nonequimolar  scenario, it is important to validate this expectation also in that case. A biased annealed system can be simulated by choosing a value different from $\frac{1}{2}$ for the parameter $q_1$ introduced above. In such a case, the fraction of particles with the orientation labeled as $1$ fluctuates around a value $x_1\neq\frac{1}{2}$. One might intuitively expect that $x_1=q_1$ after thermalization. However, we have observed that this is not the case but instead $x_1<q_1$ if $q_1>\frac{1}{2}$. This means that there are more rejections in the attempts to assign the majority orientation than in the attempts to assign the minority one.

In particular, we have  studied  biased systems with  $\lambda=1.2$,  $\rho=0.5$, and $T^*=1$, observing that the choices $q_1=0.60$, $0.70$, and $0.80$ lead to average fractions $x_1=0.5883$, $0.6789$, and $0.7748$, respectively. As an illustrative example, figure \ref{fig11} compares the four correlation functions $g_{ij}(r)$ obtained in our MC simulations with $q_1=0.80$ against  the exact solution for the quenched system with $x_1= 0.7748$. Again, an excellent agreement is apparent. The simulation points in the case of $g_{22}(r)$ are a bit noisier because in that case the  $2$--$2$ pairs represent about $5\%$ of the total number of pairs. The simulation (theoretical) values of the excess internal energy per particle are $|u_\ex|/\epsilon=0.09313$ ($0.09307$), $0.08364$ ($0.08357$), and $0.06654$ ($0.06662$) for $x_1=0.5883$, $0.6789$, and $0.7748$, respectively.

Therefore, figures \ref{fig8}--\ref{fig11} confirm that, as argued in section \ref{sec5}, the macroscopic properties of the quenched Janus fluid (where  particles have a fixed orientation) are equivalent, in the thermodynamic limit, to those of the annealed Janus fluid (where particles are allowed to flip their orientation).

\section{Summary and conclusions}
\label{sec7}

In this paper we have extensively studied the statistical--mechanical properties of one-dimensional Janus fluids. First, we have considered a general $m$-component mixture with anisotropic interactions, such that the interaction potential between a particle $\alpha$ and its nearest neighbor $\gamma=\alpha\pm 1$ depends on whether the latter is located to the left ($\gamma=\alpha-1$) or to the right ($\gamma=\alpha+1$) of $\alpha$. By carefully extending the method followed in the case of isotropic interactions \cite{S16}, we have derived the exact solution in the isothermal--isobaric ensemble. By particularizing to a binary mixture ($m=2$) with the Kern--Frenkel potential \cite{KF03}, as given by equation \eref{2},  the pair correlation functions $g_{ij}(r)$ and thermodynamic quantities (density and internal energy) are obtained as explicit functions of pressure, temperature, and composition. The  mixture  represents what we have called a \emph{quenched} Janus fluid since the orientation of the active face of each particle is kept fixed.

An interesting result is the absence of a Fisher--Widom transition between an oscillatory asymptotic decay of $h_{ij}(r)\equiv g_{ij}(r)-1$ (if the repulsive part of the interaction dominates) and a monotonic asymptotic decay (if the attractive part of the interaction dominates), in contrast to what happens in the case of the one-dimensional isotropic SW fluid \cite{FW69}. This is a consequence of the inhibition of attractive forces, as they are restricted to pairs of particles with their active faces facing each other. However, a structural crossover exists between an oscillatory decay with a wavelength smaller than twice the hard-core diameter (at high temperatures) and an oscillatory decay with a larger wavelength (at low temperatures). The phase diagram representing this structural crossover presents a `critical' point, as illustrated by figure \ref{fig4}.

In the final part of this work we have addressed the question of whether the derived exact results for the quenched Janus fluid are applicable to the case of the \emph{annealed} Janus fluid. In the latter, all the particles are identical, so that one is dealing with a one-component system in which the particles are allowed to flip their orientation and, as a consequence, the number of particles with either orientation fluctuates around a certain average value. In section \ref{sec5} we have presented compelling arguments in favor of the  quenched$\leftrightarrow$annealed equivalence in the thermodynamic limit. This has been further supported by comparison between the theoretical results for quenched systems and MC simulations for annealed systems (both unbiased and biased). Structural as well as thermodynamic quantities are seen to exhibit an excellent agreement. This in turn validates the theoretical results derived in this paper.

While most of the results presented in this paper apply to fluids confined to one-dimensional geometries, we believe that they can contribute to a better understanding of some of the peculiar physical properties of Janus fluids and also serve as a benchmark to test theoretical approaches. Additionally, the equivalence between the quenched and annealed systems gives support to the three-dimensional (quenched) up-down Janus  mixture model considered in Refs.\ \cite{MFGS13,FGMS13}.

\ack
The authors are grateful to an anonymous Referee for his/her constructive recommendations. A.S. acknowledges the financial support of the Spanish Agencia Estatal de Investigaci\'on through Grant No. FIS2016-76359-P and the Junta de Extremadura (Spain)
through Grant No. GR18079, both partially financed by Fondo Europeo de Desarrollo Regional funds.

\appendix
\section{Consistency tests
\label{appA}}

\subsection{Virial route}
In a general one-dimensional mixture (with isotropic or anisotropic interactions), the virial equation of state reads
\beq
\frac{\beta p}{\rho}=1-\rho\beta\sum_{i,j}x_ix_j\int_{0}^\infty \rmd  r\,r
g_{ij}(r) \frac{\partial \phi_{ij}(r)}{\partial r}.
\label{71}
\eeq
Now, since the interaction $\phi_{ij}(r)$ does not extend beyond the nearest neighbors, we can replace $g_{ij}(r)\to \rho_j^{-1}p_{ij}^{(1,+)}(r)=\rho^{-1}K_{ij}\rme^{-\beta pr}\rme^{-\beta\phi_{ij}(r)}$ in equation \eref{71}, so that
\beq
\frac{\beta p}{\rho}=1+\sum_{i,j}x_ix_jK_{ij}\int_{0}^\infty \rmd  r\,r
\rme^{-\beta pr}\frac{\partial \rme^{-\beta\phi_{ij}(r)}}{\partial r}.
\label{71b}
\eeq
Integrating by parts,
\beq
\frac{\beta p}{\rho}=1-\sum_{i,j}x_ix_jK_{ij}\int_{0}^\infty \rmd  r\,\left(1+p\frac{\partial}{\partial p}\right)\rme^{-\beta pr}
 \rme^{-\beta\phi_{ij}(r)}.
\label{71c}
\eeq
This equation can be rewritten as
\beq
\frac{1}{\rho}=\frac{1}{\beta p}-\sum_{i,j}x_ix_jK_{ij}\left(\Omega'_{ij}+\frac{\Omega_{ij}}{\beta p}\right)=-\sum_{i,j}x_ix_jK_{ij}\Omega'_{ij},
\label{71d}
\eeq
where in the last step we have taken into account the normalization condition \eref{norm}. Equation \eref{71d} is the generalization of equation \eref{6.34} to an arbitrary number of components.

\subsection{Compressibility route}
According to this route,
\beqa
\chi&\equiv& k_{\text{B}}T\left(\frac{\partial\rho}{\partial  p}\right)_{T,x_1}\nn
&=&\frac{\left[1+\rho x_1\widetilde{h}_{11}(0)\right]\left[1+\rho x_2\widetilde{h}_{22}(0)\right]-\rho^2 x_1 x_2\widetilde{h}_{12}(0)\widetilde{h}_{21}(0)}{1+\rho x_1 x_2\left[\widetilde{h}_{11}(0)+\widetilde{h}_{22}(0)-\widetilde{h}_{12}(0)-\widetilde{h}_{21}(0)\right]},
\label{chi}
\eeqa
where $\widetilde{h}_{ij}(\mathbf{k})=\int \rmd\mathbf{r}\, \rme^{\rmi \mathbf{k}\cdot \mathbf{r}}h_{ij}(\mathbf{r})$
is the Fourier transform of the total correlation function $h_{ij}(\mathbf{r})$, $\rmi$ being the imaginary unit. In the particular case of one-dimensional systems,
\beq
\widetilde{h}_{ij}(k_x)=\widetilde{h}_{ji}(-k_x)=\left[H_{ij}(s)+H_{ji}(-s)\right]_{s=\rmi k_x},
\label{6.17a}
\eeq
so that the zero wavenumber limit is
\beq
  \widetilde{h}_{ij}(0)=\widetilde{h}_{ji}(0)=\lim_{s\to 0}\left[G_{ij}(s)+G_{ji}(-s)\right],
  \label{6.17}
  \eeq
where equation \eref{15} has been taken into account.
Making use of equations \eref{14}, \eref{35},  and \eref{22}--\eref{26}, and after some algebra, one finds
\beq
\fl
\widetilde{h}_{11}(0)=\rho {J}-2\frac{x_2{K_{22}}\Omega_{22}'}{x_1 {K_{12}}\Omega_{12}}-\frac{2}{\rho x_1},
\quad \widetilde{h}_{22}(0)=\rho {J}-2\frac{x_1{K_{11}}\Omega_{11}'}{x_2 {K_{12}}\Omega_{12}}-\frac{2}{\rho x_2},
\label{GAA}
\eeq
\beq
\widetilde{h}_{12}(0)=\widetilde{h}_{21}(0)={\rho} {J}+\frac{\Omega_{12}'}{\Omega_{12}}+\frac{\Omega_{21}'}{\Omega_{21}},
\label{GAB}
\eeq
where
 \beq
\fl
{J}\equiv x_1^2 K_{11} \Omega_{11}'' + x_2^2 K_{22} \Omega_{22}'' +  x_1 x_2  \left(K_{12}\Omega_{12}''+K_{21}\Omega_{21}''
{ -2 K_{12}\frac{
  \Omega_{11}'\Omega_{22}' -
   \Omega_{12}'\Omega_{21}'}{\Omega_{21}}}\right).
 \label{J}
 \eeq

By inserting equations \eref{GAA}--\eref{J} into the right-hand side of  equation \eref{chi}, it can be verified that the resulting expression for the isothermal susceptibility $\chi$  indeed coincides with the one obtained as $(\partial\rho/\partial \beta p)_{\beta,x_1}$ from equation \eref{6.34}. Also, it can be checked that the denominator on the right-hand side of equation \eref{chi} reduces to
\beq
1+\rho x_1 x_2 \left[\widetilde{h}_{11}(0)+\widetilde{h}_{22}(0)-\widetilde{h}_{12}(0)-\widetilde{h}_{21}(0)\right]=\sqrt{1-4x_1x_2R}.
\label{Denomi}
\eeq
Therefore,  $\chi$ never diverges, what confirms  the classical proof \cite{vH50} by van Hove  about the absence of phase transitions  in  one-dimensional nearest-neighbor  models.

\subsection{Energy route}
In general, the excess internal energy per particle  in a one-dimensional mixture is
\beq
u_{\text{ex}}=\rho\sum_{i,j}x_ix_j\int_0^\infty \rmd  r\,
g_{ij}(r)\phi_{ij}(r).
\label{87}
\eeq
As in the case of equation \eref{71}, we can replace $g_{ij}(r)\to \rho^{-1}K_{ij}\rme^{-\beta pr}\rme^{-\beta\phi_{ij}(r)}$ in equation \eref{87}. Additionally, taking into account equation \eref{upsilon}, we obtain
\beq
u_{\text{ex}}=\sum_{i,j}x_i x_jK_{ij}\Omega_{ij}\Upsilon_{ij}.
\label{uex}
\eeq
Using the properties \eref{40}, it is straightforward to check that $u_{\text{ex}}=U/N-k_{\text{B}}T/2$, where $U$ is given by equation \eref{UN1N2}.
Note, however, that equation \eref{uex} applies to any number of components, while equation \eref{UN1N2} refers to binary mixtures only.

\section{Sticky-hard-sphere limit
\label{appB}}

In the sticky-hard-sphere (SHS) limit, the SW depth $\epsilon$ goes to infinity (so that $\theta\to\infty$) while the width $\lambda-1$ goes to zero by keeping the stickiness parameter $\tau^{-1}\equiv (\lambda-1)\theta$ fixed.
In that case, $\Omega_{12}(s)$ in equation \eref{44} becomes
\beq
\Omega_{12}(s)=\Omega(s)\left\{1-\tau^{-1}\left[1+\frac{s\Omega'(s)}{\Omega(s)}\right]\right\}=\Omega(s)\left(1+\tau^{-1}s\right).
\label{B.1}
\eeq
The general equation of state \eref{6.34} reduces to a quadratic equation for the pressure whose physical root is
\beq
\fl
\beta p=\frac{\rho}{1-\rho}\left[1-F\left(\tau\frac{1-\rho}{\rho}\right)\right], \quad
F(z)\equiv\frac{1+z-\sqrt{\left(1+z\right)^2-4x_1x_2}}{2}.
\label{B.54}
\eeq
The associated first few virial coefficients are
\beq
B_2=1-x_1x_2\tau^{-1},\quad B_3=1-x_1x_2\tau^{-1}\left(2-\tau^{-1}\right),
\label{B.2}
\eeq
\beq
B_4=1-x_1x_2\tau^{-1}\left[3-3\tau^{-1}+\tau^{-2}\left(1+x_1x_2\right)\right].
\label{B.3}
\eeq
Obviously, the same expressions are obtained by taking the SHS limit in equations \eref{B2B3} and \eref{B4}.
In the high-temperature and low-temperature limits, equation \eref{B.54} yields
\beq
\beta p=\frac{\rho}{1-\rho}\left[1-\frac{\rho}{1-\rho}x_1x_2\tau^{-1}\right]+\mathcal{O}(\tau^{-2}),
\label{B.4}
\eeq
\beq
\lim_{\tau\to 0}\beta p=\max(x_1,x_2)\frac{\rho}{1-\rho},
\label{B.5}
\eeq
where in equation \eref{B.5} we have taken into account that $1-\sqrt{1-4x_1x_2}=2\min(x_1,x_2)$.
As expected, equations \eref{B.4} and \eref{B.5} are fully consistent with equations \eref{betapHT} and \eref{betapLT}, respectively.

In terms of density, the amplitudes \eref{40} and \eref{6.33} become
\beq
\fl
K_{11}=\frac{\rme^{\beta p}}{x_1}\left[\beta p-\frac{\tau+\beta p}{x_1}F\left(\tau\frac{1-\rho}{\rho}\right)\right],\quad
K_{22}=\frac{\rme^{\beta p}}{x_2}\left[\beta p-\frac{\tau+\beta p}{x_2}F\left(\tau\frac{1-\rho}{\rho}\right)\right],
\label{B.52}
\eeq
\beq
K_{12}=\frac{\tau \rme^{\beta p}}{x_1x_2} F\left(\tau\frac{1-\rho}{\rho}\right),
\quad
K_{21}=\frac{(\tau+\beta p)\rme^{\beta p}}{x_1x_2}F\left(\tau\frac{1-\rho}{\rho}\right).
\label{B.49}
\eeq
As a consequence, equation \eref{uex1} simply reduces to
\beq
\frac{u_\ex}{\epsilon}=-F\left(\tau\frac{1-\rho}{\rho}\right).
\label{B.50}
\eeq
Therefore,
\beqa
\fl
u_2=-x_1x_2\tau^{-1},\quad u_3=-x_1x_2\tau^{-1}(1-\tau^{-1}),\quad
u_4=-x_1x_2\tau^{-1}\left[(1-\tau^{-1})^2+x_1x_2\tau^{-2}\right],\nn
\eeqa
\beq
\frac{u_\ex}{\epsilon}=-x_1x_2\frac{\rho}{1-\rho}\tau^{-1}+\mathcal{O}(\tau^{-2}),\quad
\lim_{\tau\to 0}\frac{u_\ex}{\epsilon}=-\min(x_1,x_2).
\label{uexLT_SHS}
\eeq

In what concerns the structural properties, we note that in the SHS limit
\beq
\Psi^{(k_1,k_2)}(s)\to \bar{\Psi}^{(k_1+k_2,0)}(s),
\eeq
\beq
\X\left[\Psi^{(k_1,k_2)}(s)-\lambda \Psi^{(k_1-1,k_2+1)}(s)\right]\to \tau^{-1}\bar{\Psi}^{(k_1+k_2-1,1)}(s),
\eeq
where
\beq
\bar{\Psi}^{(k_1,k_2)}(s)\equiv \frac{\left[\Omega(s+\beta p)\right]^{k_1}\rme^{-k_2\left(s+\beta p\right)}}{D(s)}.
\label{B.Psi}
\eeq
As a consequence, equations \eref{57}--\eref{60} become
\beq
G_{11}(s)=\frac{K_{11}}{\rho}\bar{\Psi}^{(1,0)}(s)+\frac{x_2K_{11}K_{22}\tau^{-1}}{\rho}
\bar{\Psi}^{(1,1)}(s),
\label{B.57}
\eeq
\beq
G_{22}(s)=\frac{K_{22}}{\rho}\bar{\Psi}^{(1,0)}(s)+\frac{x_1K_{11}K_{22}\tau^{-1}}{\rho}
\bar{\Psi}^{(1,1)}(s),
\label{B.58}
\eeq
\beq
G_{12}(s)=\frac{K_{12}}{\rho}\bar{\Psi}^{(1,0)}(s)+\frac{K_{12}\tau^{-1}}{\rho}\bar{\Psi}^{(0,1)}(s),
\label{B.59}
\eeq
\beq
G_{21}(s)=\frac{K_{21}}{\rho}\bar{\Psi}^{(1,0)}(s),
\label{B.60}
\eeq
The determinant $D(s)$ can be written in this case as
\beq
D(s)=1-a\Omega(s+\beta p)-\bar{b}\Omega(s+\beta p)\rme^{-(s+\beta p)},
\eeq
where $a$ is still given by equation \eref{ab} and $\bar{b}\equiv x_1x_2 K_{11}K_{22}\tau^{-1}$.
Using the mathematical identity
\beq
\left(1-ax- \bar{b} xy\right)^{-1}=\sum_{n=0}^\infty\sum_{\ell=0}^{n}\bar{C}_{n,\ell}
x^ny^\ell,\quad \bar{C}_{n,\ell}\equiv \frac{n!}{\ell!(n-\ell)!}
a^{n-\ell}\bar{b}^\ell,
\label{B.62}
\eeq
we have
\beq
\bar{\Psi}^{(k_1,k_2)}(s)=\sum_{n=0}^\infty\sum_{\ell=0}^{n}\bar{C}_{n,\ell}
\frac{\rme^{-(n+\ell+k_1+k_2)(s+\beta p)}}{(s+\beta p)^{n+k_1}}.
\eeq
Thus, the Laplace property \eref{65} allows us to write the inverse Laplace transform of $\bar{\Psi}^{(k_1,k_2)}(s)$ as
\beq
\fl
\bar{\psi}^{(k_1,k_2)}(r)=\rme^{-\beta p r}\sum_{n=0}^\infty\sum_{\ell=0}^{n}\bar{C}_{n,\ell}\frac{(r-n-\ell-k_1-k_2)^{n+k_1-1}}{(n+k_1-1)!}\Theta(r-n-\ell-k_1-k_2).
\label{B.psi}
\eeq
This expression holds if $k_1>0$. On the other hand, if $k_1=0$,
\beq
\fl
\bar{\psi}^{(0,k_2)}(r)=\rme^{-\beta p r}\left[\delta(r-k_2)+\sum_{n=1}^\infty\sum_{\ell=0}^{n}\bar{C}_{n,\ell}\frac{(r-n-\ell-k_2)^{n-1}}{(n-1)!}\Theta(r-n-\ell-k_2)\right],
\label{B.psi0}
\eeq
where use has been made of $\mathcal{L}^{-1}\left[{\rme^{-n(s+\beta p)}}\right]= \rme^{-\beta p r}\delta(r-n)$.

\section*{References}

\providecommand{\newblock}{}

\end{document}